\documentclass[aps,prl,reprint,amsmath,amssymb,superscriptaddress]{revtex4-2}
\usepackage{booktabs}
\usepackage{multirow}
\usepackage{amsmath}
\usepackage{bm}
\usepackage{graphicx}

\newcommand{\msun}{M_\odot}
\newcommand{\kpc}{\mathrm{kpc}}
\newcommand{\gyr}{\mathrm{Gyr}}
\begin{document}

\title{Baryon-Enhanced Strong Lensing Signatures of Self-Interacting Dark Matter Subhalos}

\author{Meiwen Yang}
\affiliation{Purple Mountain Observatory, Chinese Academy of Sciences, Nanjing 210033, China}
\affiliation{School of Astronomy and Space Sciences, University of Science and Technology of China, Hefei 230026, China}

\author{Daneng Yang}
\email{yangdn@pmo.ac.cn}
\affiliation{Purple Mountain Observatory, Chinese Academy of Sciences, Nanjing 210033, China}
\affiliation{School of Astronomy and Space Sciences, University of Science and Technology of China, Hefei 230026, China}

\author{Siyuan Hou}
\affiliation{Purple Mountain Observatory, Chinese Academy of Sciences, Nanjing 210033, China}
\affiliation{School of Astronomy and Space Sciences, University of Science and Technology of China, Hefei 230026, China}

\author{Yi-Ying Wang}
\affiliation{Purple Mountain Observatory, Chinese Academy of Sciences, Nanjing 210033, China}

\author{Yiping Shu}
\affiliation{Purple Mountain Observatory, Chinese Academy of Sciences, Nanjing 210033, China}
\affiliation{School of Astronomy and Space Sciences, University of Science and Technology of China, Hefei 230026, China}

\author{Yue-Lin Sming Tsai}
\affiliation{Purple Mountain Observatory, Chinese Academy of Sciences, Nanjing 210033, China}
\affiliation{School of Astronomy and Space Sciences, University of Science and Technology of China, Hefei 230026, China}

\date{\today}

\begin{abstract}
Self-interacting dark matter (SIDM) is generally expected to suppress subhalo strong lensing through core formation, with subsequent gravothermal collapse potentially enhancing it. We show that compact stars can make a viable population of galaxy-hosting SIDM subhalos more efficient lenses than their cold dark matter (CDM) counterparts. When the baryonic potential weakens or reverses the inner velocity-dispersion gradient, SIDM halos suppress or bypass core formation, allowing inner densities to grow from early times. The evolved SIDM subhalos produce larger deflection angles and tangential caustics than in CDM. Such galaxy-hosting subhalos also dominate the galaxy-galaxy strong-lensing (GGSL) budget in clusters, yet lie far below the resolution of cosmological simulations. We calibrate a baryonic parametric SIDM model to predict their effects in the three most massive TNG300-1 clusters, finding GGSL cross sections 1.4--2.7 times those in CDM, comparable to the excess inferred in recent analyses. Baryon-enhanced SIDM lensing can thus account for the observed GGSL excess and predicts stronger perturbations from luminous satellites, testable through resolved distortions of background arcs and rings. 
\end{abstract}

\maketitle

\section{Introduction}
\label{sec:introduction}

In self-interacting dark matter (SIDM), gravothermal collapse can produce subhalos that are denser and more efficient gravitational lenses than their collisionless counterparts \citep{Balberg:2002ue,Tulin:2017ara,Adhikari:2022sbh}.
Such structures can be probed through gravitational imaging, which detects perturbations to lensed arcs and rings \citep{Vegetti:2009gravitational,Vegetti:2012mc,Minor:2020hic,Nadler:2023nrd,Yu:2025tmp,Enzi:2024ygw,Zhang:2026gur,Li:2025kpb,He:2025wco,Tekverk:2026slk}, and galaxy-galaxy strong lensing (GGSL) searches, which identify additional lensing configurations generated by cluster substructures \citep{Limousin:2006cv,Meneghetti:2020yif}.
Quasar flux ratios and the spatial distribution of highly magnified stars also test SIDM core collapse~\cite{Gilman:2026qgr,Li:2026abn}.
Cluster lens models revealed a GGSL probability more than an order of magnitude above hydrodynamical cold dark matter (CDM) predictions \citep{Meneghetti:2020yif}, motivating SIDM core-collapse interpretations \citep{Yang:2021kdf}.
Subsequent studies examined its sensitivity to numerical resolution, galaxy formation, lens modeling, and sample selection, finding a reduced, methodology-dependent excess of up to a factor of about $2$--$3$, with its magnitude remaining under debate~\citep{Robertson:2021ggsl,Bahe:2021bcs,Meneghetti:2023fug,Granata:2023ryu,Tokayer:2024wwo,Roche:2026jan}. A core-collapse explanation requires a substantial population of subhalos with enhanced projected masses on strong-lensing scales.

To produce this population, halos must collapse quickly while retaining enough projected mass to lens. Dark-matter-only SIDM halos usually form low-density cores first, which weakens lensing and makes them more vulnerable to tides. Tidal stripping can accelerate core collapse~\citep{Nishikawa:2019lsc,Yang:2023jwn}, but it also causes mass loss that can lower the projected mass needed for lensing \citep{Yang:2021kdf,Zeng:2021ldo,Kong:2025kkt}. Scattering with host particles can further evaporate subhalos and slow down collapse, favoring cross sections that decrease from subhalo to cluster velocities \citep{Zeng:2021ldo,Klemmer:2026ivv}. The strong-lensing response of SIDM therefore depends on the balance between gravothermal contraction and mass loss.

Baryons can significantly alter this story. Compact stellar components deepen the inner potential and contract the dark matter distribution, accelerating SIDM gravothermal evolution at a rate that depends on their mass fraction and compactness~\citep{Sameie:2018chj,Despali:2019sidm,Jiang:2022aqw,Rose:2022sidm,Zhong:2023yzk,Mezghanni:2026vnn}. In particular, a sufficiently deep and extended baryonic potential can suppress or bypass core formation, allowing the inner SIDM density to grow from early times \citep{Yang:2024tba,Zeng:2026umo}.
Hydrodynamical cosmological simulations have found a consistent pattern in baryon-dominated $\sim10^{12}~\msun$ halos, some of which can become denser than their CDM counterparts \citep{Correa:2024vgl,Despali:2026aida}. These findings motivate quantifying how baryon-modified gravothermal evolution changes the lensing efficiency of individual satellites and the collective signal of cluster subhalos.

In this Letter, we connect baryon-modified gravothermal evolution to the reported GGSL excess and predict enhanced lensing perturbations from luminous satellites. We identify TNG300-1 subhalos in the regime of suppressed core formation and follow six benchmarks through pre-infall evolution and subsequent tidal evolution with live stars and dark matter. We jointly calibrate a baryonic SIDM model against these and earlier simulations and predict GGSL for the three most massive TNG300-1 clusters. These predictions connect stellar structure to subhalo lensing efficiency, providing an observational test of baryon-driven gravothermal evolution.

\section{Baryonic Regimes of SIDM Subhalos}
\label{sec:regimes}

\begin{figure}[thbp]
\centering
\includegraphics[width=\columnwidth]{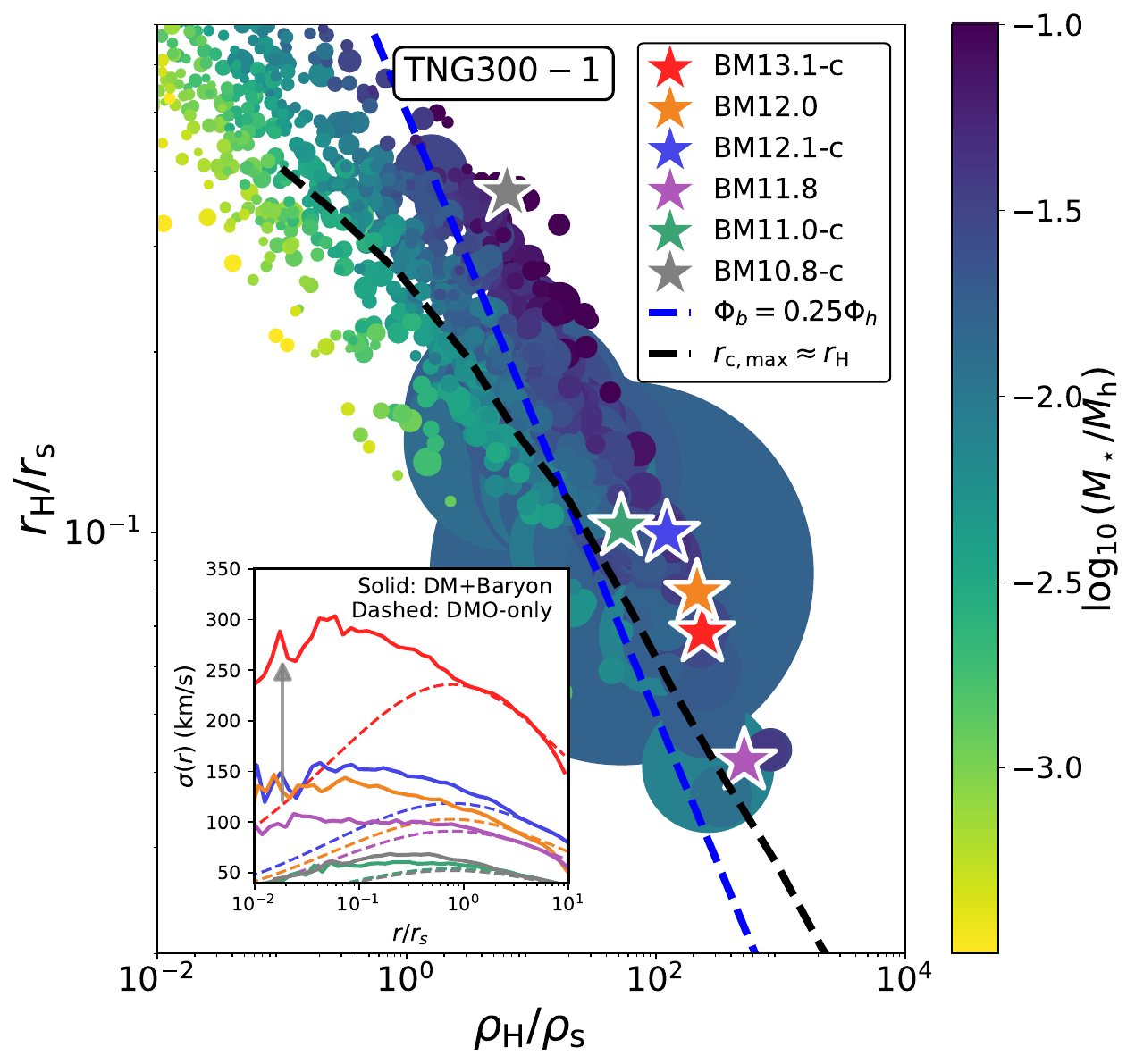}
\caption{
Subhalos of the most massive TNG300-1 cluster in the plane of Hernquist-to-NFW scale-density and scale-radius ratios, $\rho_H/\rho_s$ and $r_H/r_s$. Halo and stellar parameters are evaluated at each subhalo's dark matter peak-mass snapshot. Color indicates the stellar-to-halo mass ratio, $M_\star/M_h$, while marker size scales with halo virial radius. Stars denote the six benchmarks listed in Table~\ref{tab:lensing}. The blue dashed line marks $\Phi_b=0.25\Phi_h$, where $\Phi_b\equiv G\rho_H r_H^2$ and $\Phi_h\equiv G\rho_s r_s^2$ characterize the depths of the baryonic and NFW potentials, respectively. 
The black dashed curve marks $r_H=r_{\mathrm{c,max}}$, with $r_{\mathrm{c,max}}$ the maximum core radius predicted by the baryonic PSIDM model~\citep{Yang:2024tba}. 
A substantial fraction of the massive subhalos, including the six benchmarks, lie to the right of both curves, where baryonic effects are expected to be strong.
The inset shows the initial dark matter velocity-dispersion profiles of the six benchmarks, with (solid) and without (dashed) stars, in colors matching the main panel. The stellar potentials raise the inner dark matter velocity dispersions and weaken or reverse their radial gradients.
}
\label{fig:selection}
\end{figure}

We start by examining the subhalo population of the most massive TNG300-1 cluster at $z=0$~\citep{Nelson:2017cxy,Pillepich:2017fcc}. For each subhalo, we evaluate both halo and stellar parameters at its own dark matter peak-mass snapshot, so these configurations span different redshifts. This population serves as the physical context for our controlled benchmarks. 

Figure~\ref{fig:selection} shows these subhalos in the plane of Hernquist-to-NFW  scale-density and scale-radius ratios \citep{Hernquist:1990be,Navarro:1996gj}, $\rho_H/\rho_s$ and $r_H/r_s$. We describe the stellar and halo potential depths using $\Phi_b\equiv G\rho_Hr_H^2$ and $\Phi_h\equiv G\rho_sr_s^2$. The blue line, $\Phi_b=0.25\Phi_h$, helps identify halos where stars can significantly change the inner dark matter velocity dispersion. The contour $r_{\mathrm{c,max}}=r_H$, predicted by the baryonic parametric SIDM (PSIDM) model, separates configurations where the stellar potential confines the maximum core within $r_H$ from those where the core can still expand beyond the stellar scale despite baryonic suppression~\citep{Yang:2024tba}. Among the 504 subhalos with peak masses above $10^{11}\,\msun$, $43\%$ have $\Phi_b > 0.25\Phi_h$ and $75\%$ have $r_{\mathrm{c,max}} < r_H$, indicating that such stellar configurations are common among massive cluster members.

\begin{table*}[t]
\caption{Initial benchmark parameters and lensing results after tidal evolution. Here, $r_H$ is the Hernquist scale radius and $\sigma_T$ is the source-plane tangential caustic area. Lensing calculations use $z_l=0.439$, $z_s=3$, and a projected distance of $350\,\kpc$ from the cluster center. The benchmarks are simulated with the constant SIDM cross sections $\sigma_{\rm eff}/m$ listed here.}
\label{tab:lensing}
\centering
\scriptsize
\setlength{\tabcolsep}{4pt}
\begin{ruledtabular}
\begin{tabular}{lcccccccc}
Model & $M_h$ & $M_\star$ & $r_s$ & $r_H/r_s$ &
$\sigma_{\rm eff}/m$ &
$\theta_{\rm E,eff}^{\rm CDM}$ &
$\theta_{\rm E,eff}^{\rm SIDM}$ &
$\sigma_T^{\rm SIDM}/\sigma_T^{\rm CDM}$ \\
& ($M_\odot$) & ($M_\odot$) & ($\kpc$) &
& ($\mathrm{cm^2/g}$) &
(arcsec) & (arcsec) & \\
\hline
BM13.1-c & $1.3\times10^{13}$ & $2.5\times10^{11}$ & 50.4 & 0.07 & 1.65 & 1.52  & 2.08  & 1.89 \\
BM12.0   & $1.0\times10^{12}$ & $3.4\times10^{10}$ & 24.0 & 0.08 & 5.0  & 0.38  & 0.42  & 1.3  \\
BM12.1-c & $1.3\times10^{12}$ & $3.7\times10^{10}$ & 17.5 & 0.10 & 4.5  & 0.51  & 0.69  & 1.91 \\
BM11.8   & $5.7\times10^{11}$ & $6.0\times10^{9}$  & 15.6 & 0.04 & 6.5  & 0.2   & 0.26  & 1.7  \\
BM11.0-c & $1.0\times10^{11}$ & $1.4\times10^{9}$  & 7.0 & 0.10 & 9.0  & 0.057 & 0.12  & 5.0  \\
BM10.8-c & $6.1\times10^{10}$ & $4.1\times10^{9}$  & 3.8 & 0.36 & 8.0  & 0.09  & 0.145 & 3.2  \\
\end{tabular}
\end{ruledtabular}
\end{table*}

These comparisons motivate six benchmarks covering halo mass, stellar compactness, and concentration, shown as stars in Fig.~\ref{fig:selection} and listed in Table~\ref{tab:lensing}. All of them are on the strong-baryon side of both contours. The halo concentrations of BM13.1-c, BM12.1-c, BM11.0-c, and BM10.8-c are higher than the median concentration-mass relation by $1.5\sigma$, $1.9\sigma$, $1.1\sigma$, and $2.6\sigma$, respectively~\cite{Dutton:2014xda}. We use controlled simulations to test how compact baryons reshape their strong-lensing signatures. 

The inset of Fig.~\ref{fig:selection} presents the initial dark matter velocity-dispersion profiles. Stellar potentials increase the inner dispersions, which reduces the inward heat flow responsible for core expansion or reverses it to favor contraction~\citep{Zeng:2026umo}. We use controlled simulations to test the resulting gravothermal evolution and strong-lensing signatures.

\section{Gravothermal Evolution and Lensing Enhancement}
\label{sec:evolution}

We test these benchmarks under CDM and SIDM with live dark matter and stars. Initial conditions are generated with \texttt{SpherIC} \citep{Garrison-Kimmel:2013yys} and evolved using \texttt{GADGET-2}~\cite{Springel:2005mi} with the SIDM module developed in Refs.~\cite{Yang:2022hkm,Yang:2022mxl}. The benchmarks contain $N=1.5\times10^6$--$4\times10^6$ dark matter particles and $10^4$--$6\times10^4$ stellar particles. Both components use the same gravitational softening, $\epsilon=2R_{\rm vir}/\sqrt{N}$, with $R_{\rm vir}$ the initial halo virial radius.

We characterize the scattering strength using an effective constant cross section $\sigma_{\rm eff}/m$, calculated by averaging the velocity- and angle-dependent differential cross section with a $v^5\sin^2\theta$ weight~\citep{Yang:2022hkm,Yang:2022zkd}. Our simulations use the constant values provided in Table~\ref{tab:lensing}. For the velocity-dependent models discussed below, we adopt a Rutherford-like form
\begin{equation}
\frac{\mathrm{d}\sigma}{\mathrm{d}\cos\theta}=\frac{\sigma_0 w^4}{2\left[w^2+v^2\sin^2(\theta/2)\right]^2},
\label{eq:dsigma} 
\end{equation}
where $v$ is the relative particle velocity, $\theta$ is the scattering angle, and $\sigma_0$ and $w$ set the scattering normalization and characteristic velocity.

\begin{figure}[!htbp]
\centering
\includegraphics[width=\columnwidth]{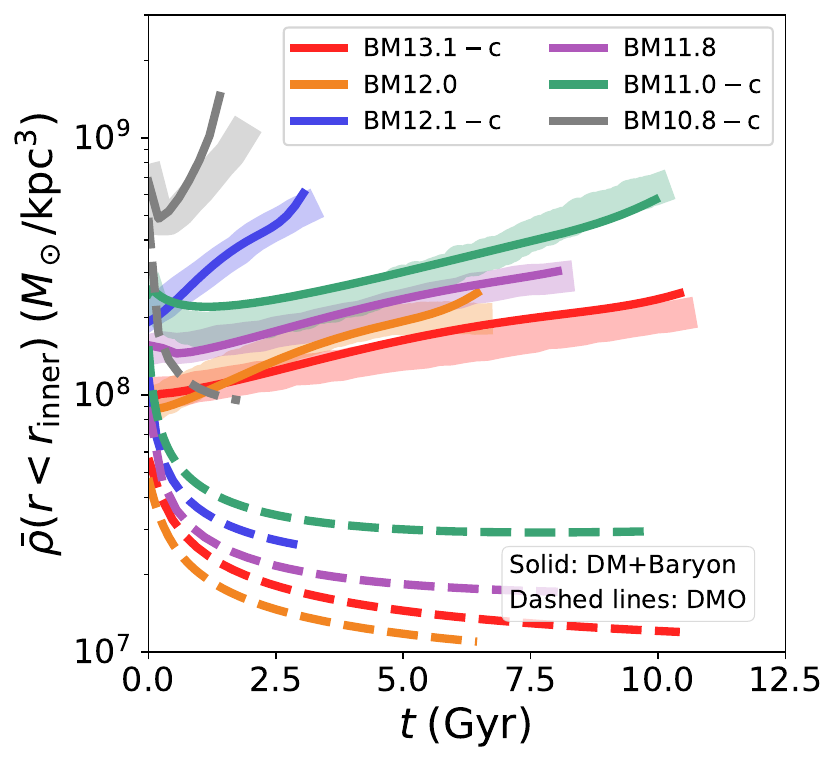}
\caption{Evolution of the averaged dark matter density within $0.1r_s$ for the benchmarks in Table~\ref{tab:lensing}, with and without stellar components. Shaded bands show controlled $N$-body results with live stellar components. Solid curves show predictions from the recalibrated baryonic parametric SIDM model, while dashed curves show dark-matter-only predictions from the original parametric model~\citep{Yang:2023jwn}. Baryons suppress core formation in all cases, and most benchmarks grow denser from early times.}
\label{fig:evolution}
\end{figure}

The gravothermal phase $\tau=t/t_{cb}$ tracks how much time $t$ has passed compared to the baryon-modified collapse timescale $t_{cb}$~\citep{Outmezguine:2022bhq,Yang:2024tba,Yang:2023jwn}. When $\tau$ approaches one, the core contracts and its central density rises quickly. Figure~\ref{fig:evolution} shows the mean dark matter densities within $0.1r_s$, with $r_s$ set to its initial NFW value. 
The presence of live stars suppresses the density decrease predicted in the dark-matter-only scenario~\cite{Yang:2023jwn,Yang:2024uqb}. Models BM13.1-c, BM12.0, BM12.1-c, and BM11.8 exhibit increasing central densities with minimal or no extended core-forming phase. In these simulations, core formation is suppressed as the initial dispersion slope at $0.05r_s$ approaches zero and is bypassed when it becomes negative. Models BM11.0-c and BM10.8-c maintain positive inner gradients and show shallow and more pronounced initial density drops, respectively, before contraction drives rapid density growth.
The live stars also respond to the evolving halo, with density changes below about $20\%$ at $0.1r_s$. While our simulations do not include stellar feedback, controlled studies of high-concentration SIDM halos have found that gravothermal collapse remains robust under strong feedback~\citep{Kong:2026piq}.

Next, we test the enhancement under cluster tides. Each benchmark enters a fixed NFW host potential at $400\,\kpc$ with a tangential velocity between $1000$ and $1400\,\mathrm{km/s}$, evolving for $2\,\gyr$ and completing about two to three pericentric passages. The host parameters, $\rho_{s,\mathrm{host}}=1.3\times10^6\,\msun\,/\mathrm{kpc}^{3}$ and $r_{s,\mathrm{host}}=438\,\kpc$, represent the most massive TNG300-1 cluster at $z=0$. These idealized orbits yield $r_t/r_s=2.0$--$5.7$, extending into the more strongly truncated tail of the population at $z\simeq0.44$, where $90\%$ of subhalos above $10^{10}\,\msun$ have $r_t/r_s>3.5$. 

After tidal evolution, we calculate lensing maps from the paired CDM and SIDM outputs, including both dark matter and stars. 
Both calculations use the same host convergence and shear, a projected distance of $350\,\kpc$ from the cluster center, and source redshift $z_s=3$. We adopt the lens redshift of MACS J1206.2$-$0847, $z_l=0.439$~\citep{Meneghetti:2023fug}.
We define the effective Einstein radius of a tangential critical curve as
$\theta_{\rm E,eff}=\sqrt{A_{\rm crit}/\pi}$, where $A_{\rm crit}$ is the area enclosed on the image plane. The area inside the corresponding source-plane tangential caustic, $\sigma_T$, gives the subhalo strong-lensing cross section.

All six evolved SIDM benchmarks yield larger effective Einstein radii and tangential caustic areas than their matched CDM counterparts, as summarized in Table~\ref{tab:lensing}. The caustic areas increase by factors of $1.3$--$5$, with the largest enhancements in BM11.0-c and BM10.8-c reaching $5$ and $3.2$, respectively. 
Earlier simulations with strong truncation, $r_t/r_s<1$, found primarily radial caustic enhancement~\cite{Yang:2021kdf}. Our benchmarks instead show enhanced tangential caustics under less severe stripping, more typical of cluster subhalos.

To extend these results to a cluster population, we recalibrate the baryonic parametric SIDM model using our six live-stellar benchmarks and the simulations from Ref.~\cite{Yang:2024tba}, which span diverse stellar mass fractions and compactness. The solid curves in Fig.~\ref{fig:evolution} reproduce the overall isolated density evolution, including suppressed core formation and subsequent contraction. The integral approach accumulates SIDM-induced structural changes along CDM evolution histories~\cite{Yang:2023jwn,Yang:2024uqb} and also reproduces the tidally evolved dark matter profiles. These comparisons test the model in both isolated and tidally evolving stellar systems before using it for the full population.

\section{Luminous Strong-Lensing Perturbers}
\label{sec:perturbers}

\begin{figure}[!t]
\centering
\includegraphics[width=\columnwidth]{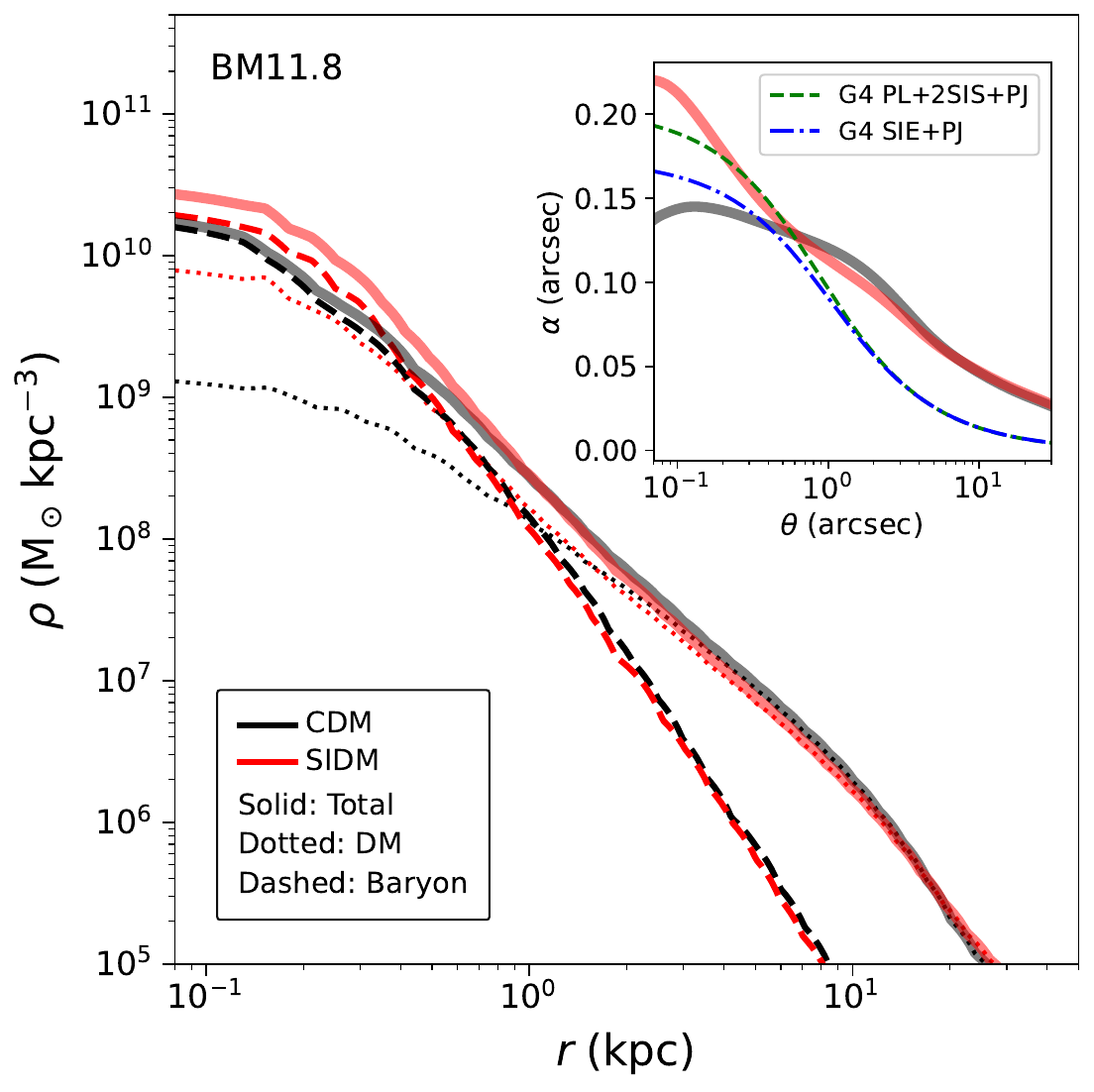}
\caption{Enhanced lensing by a galaxy-hosting SIDM subhalo, illustrated with the BM11.8 benchmark after tidal evolution. The main panel shows density profiles for CDM (black) and SIDM (red), with solid, dotted, and dashed curves denoting total matter, dark matter, and stars, respectively. The inset shows the increase in total deflection angles near the satellite center caused by SIDM contraction. For scale, green dashed and blue dash-dotted curves show the published PL+2SIS+PJ and SIE+PJ lens-model predictions for G4, an observed luminous satellite with a comparable stellar mass \citep{Vegetti:2010wa}.}
 \label{fig:perturber}
\end{figure}

Our benchmark results predict enhanced inner deflections from galaxy-hosting SIDM satellites, which can strengthen distortions of background arcs and rings. The luminous satellite G4 in SDSS J120602.09+514229.5 shows an observed configuration where this prediction can be tested \citep{Vegetti:2010wa}. 

Figure~\ref{fig:perturber} illustrates the SIDM response of our $M_h=5.7\times10^{11}\,\msun$ benchmark, whose stellar mass of $6\times10^9\,\msun$ is comparable to that inferred for G4. Using $r_t\propto d[m/M_{\rm host}]^{1/3}$, the benchmark at $320\,\kpc$ in a $10^{15}\,\msun$ cluster has approximately equivalent tidal conditions at $d\simeq69\,\kpc$ in a $10^{13}\,\msun$ group, compatible with G4's projected separation of $22.5\,\kpc$~\citep{Vegetti:2010wa}. We calculate the total deflection angles, including stars, at G4's lens and source redshifts, $z_l=0.422$ and $z_s=2.001$. After tidal evolution, its higher inner SIDM density increases the deflection angles near the satellite center, including the stellar contribution. At these small separations, the SIDM prediction approaches the values inferred from the published G4 lens models.

For low-mass, predominantly dark subhalos, SIDM core collapse can make otherwise undetectable perturbations measurable, increasing the number of detectable perturbers \citep{Minor:2020hic,Tekverk:2026slk,Li:2015xpc,Vegetti:2012mc}. Massive, galaxy-hosting satellites can already produce measurable perturbations in CDM, so their occurrence depends mainly on their abundance and alignment with background images. SIDM contraction primarily strengthens these existing perturbations. 
Their stellar light independently locates the satellites, making the amplitude and radial structure of the perturbation a direct target for SIDM tests. The characteristic prediction is a stronger perturbation at a known galaxy position, extending SIDM searches beyond changes in the detectable abundance of low-mass dark perturbers. 

\section{Strong Lensing by Cluster Subhalos}
\label{sec:lensing}

\begin{figure}[t]
\centering
\includegraphics[width=\columnwidth]{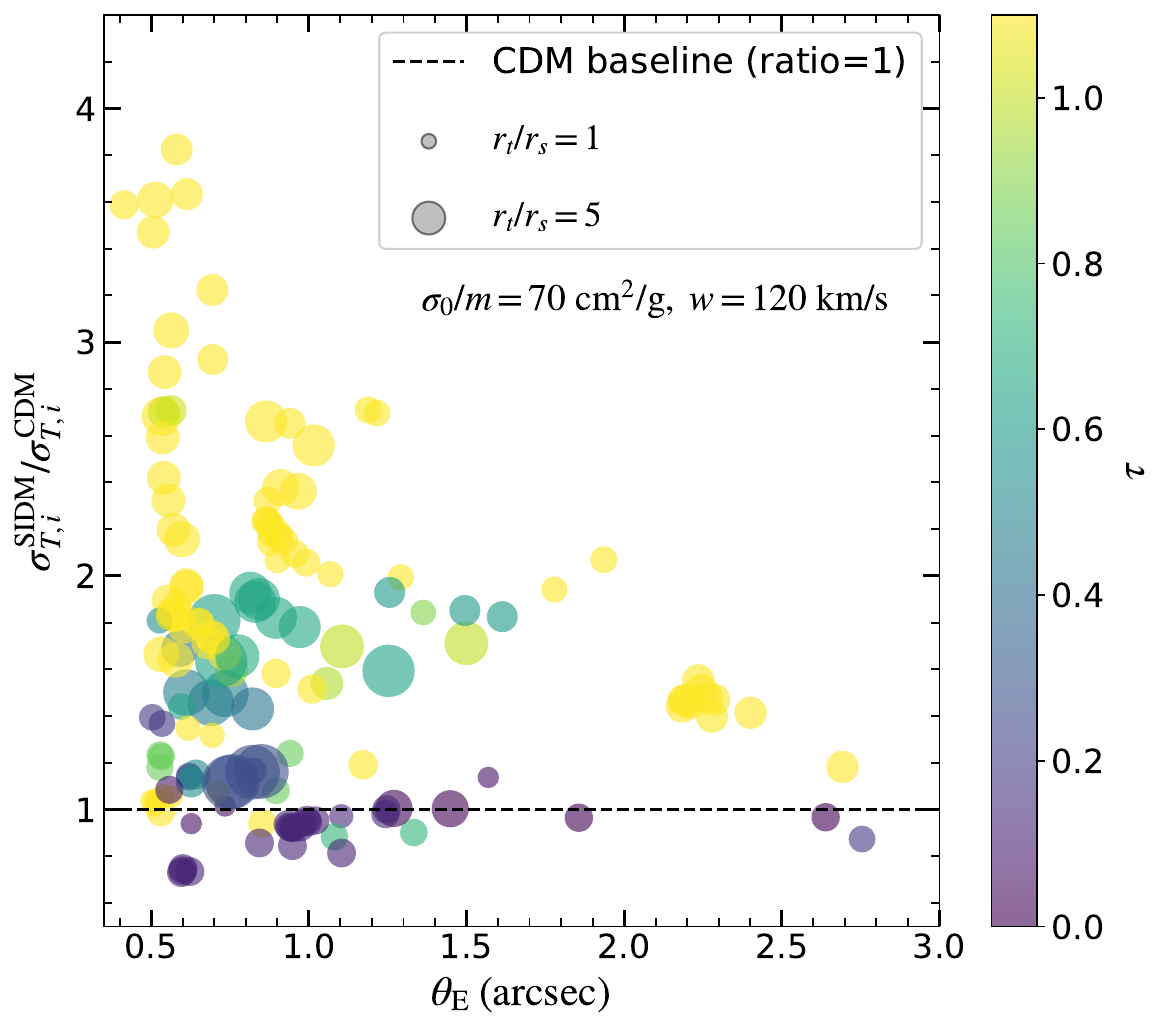}
\caption{Individual subhalo lensing responses underlying the population-level GGSL enhancement. Each point represents a matched CDM/SIDM subhalo lens, with samples pooled from all 11 viewing directions for each of the three most massive TNG300-1 clusters, plotted against the SIDM effective Einstein radius. The vertical axis shows the ratio of its source-plane tangential caustic areas, $\sigma_{T,i}^{\rm SIDM}/\sigma_{T,i}^{\rm CDM}$. Colors indicate the gravothermal phase $\tau$, marker sizes encode $r_t/r_s$, and the dashed line marks equal cross sections. 
Most subhalo lenses show enhanced caustic areas relative to CDM, with stronger enhancement generally associated with larger $\tau$. SIDM profiles are predicted using the recalibrated baryonic parametric model with $\sigma_0/m=70\,\mathrm{cm^2/g}$ and $w=120\,\mathrm{km/s}$. The calculations use identical stellar distributions and lensing environments, with $z_l=0.439$ and $z_s=3$. }
\label{fig:population}
\end{figure}

GGSL tests the collective lensing response of cluster members on galaxy scales~\citep{Meneghetti:2020yif,Meneghetti:2022apr,Meneghetti:2023fug}. We apply the recalibrated model to the resolved subhalo populations of the three most massive TNG300-1 clusters at the snapshot nearest $z_l=0.439$, with host masses spanning $5.32\times10^{14}$--$1.36\times10^{15}\,\msun$. The integral approach incorporates accretion through CDM subhalo histories. We take the reference NFW parameters from the dark-matter-only simulation and the stellar parameters from matched full-physics halos to model baryonic contraction and SIDM gravothermal evolution. The stellar distributions are kept identical in CDM and SIDM, consistent with their small differences in the controlled simulations. The paired calculations share the lensing geometry and host potential, with suppressed scattering at cluster velocities motivating the neglect of host core formation and subhalo evaporation.

For each cluster, we analyze 11 viewing directions, comprising the original orientation and five rotations in $18^\circ$ increments about each of two orthogonal axes. In each view, we sum the source-plane tangential caustic areas of subhalo lenses with $0.5''<\theta_{\rm E,eff}\leq3''$ within a $200''\times200''$ field to obtain $\sigma_{\rm GGSL}$. We then average these cross sections for CDM and SIDM separately and report $\langle\sigma_{\rm GGSL}^{\rm SIDM}\rangle/\langle\sigma_{\rm GGSL}^{\rm CDM}\rangle$.

Figure~\ref{fig:population} shows that most subhalo lenses have larger tangential caustic areas than in CDM, with enhancement generally increasing with $\tau$. Halo masses, concentrations, and baryonic configurations contribute substantial scatter. Larger $\tau$ need not always imply stronger lensing, as also found during advanced collapse in isolated baryon-free halos~\citep{vandenBosch:2026apa}. Most contributing lenses have $r_t\gg r_s$, suggesting limited direct tidal truncation of their inner regions.

We consider two velocity-dependent SIDM models. The first has $\sigma_0/m=20\,\mathrm{cm^2/g}$ and $w=500\,\mathrm{km/s}$, giving a nearly constant cross section at dwarf-halo velocities. The second has $\sigma_0/m=70\,\mathrm{cm^2/g}$ and $w=120\,\mathrm{km/s}$~\citep{Nadler:2025jwh}. Their mean GGSL cross-section ratios range from $1.43$--$2.34$ and $1.45$--$2.65$, respectively, among the three clusters. Both models yield enhancements comparable to recent estimates of the residual excess~\citep{Roche:2026jan}. Our matched calculations fix the cluster sample and lensing environment, helping isolate the SIDM enhancement from sample-selection biases.

Existing hydrodynamical cluster simulations have investigated both velocity-independent~\citep{Robertson:2018anx,Harvey:2025paa,Lee:2026otu} and velocity-dependent SIDM~\citep{Robertson:2018anx,Ragagnin:2024deh}. Resolving the evolution of cluster dwarfs requires substantially higher resolution. A reference target of $10^5$ particles per $10^{11}\,\msun$ dark matter subhalo~\citep{Mace:2024uze} requires mass-resolution improvements of approximately $10^3$ and 60 over Dianoga SIDM and TNGC-SIDM, respectively~\citep{Ragagnin:2024deh,Lee:2026otu}.

\section{Discussion and Conclusions}
\label{sec:discussion}

Our results demonstrate that compact stars can make a viable population of SIDM subhalos stronger gravitational lenses than their CDM counterparts.
The required stellar configurations appear in TNG300-1, and controlled simulations show enhanced deflection angles and tangential caustics after isolated evolution followed by tidal evolution.
The connection between stellar structure and gravothermal evolution extends core-collapse interpretations of GGSL beyond strongly stripped subhalos and offers SIDM tests through enhanced perturbations by luminous satellites. Mass segregation in two-component SIDM can further enhance inner densities, potentially accounting for more extreme GGSL efficiencies \citep{Yang:2025xsp}. 

Improved simulations and lens modeling will clarify the magnitude of the GGSL excess and the contribution of baryon-modified SIDM evolution \citep{Roche:2026jan}. 
The predicted enhancement of luminous-satellite deflections offers a test beyond the total GGSL rate, through joint modeling of stellar light and resolved image distortions \citep{Vegetti:2010wa,He:2025wco}.
Larger cluster-lensing samples, high-resolution imaging, and stellar-kinematic measurements can test the predicted enhancement and its dependence on stellar structure \citep{Euclid:2026ulx,Euclid:2025oji,Cao:2023bnl,Zhao:2024alp,JUSTTeam:2024wce,Bergamini:2023gbq}. Baryonic structure can thus determine the sign of SIDM's strong-lensing response and inform the interpretation of subhalo lens populations.

The recalibrated baryonic PSIDM model is publicly available at \url{https://github.com/MeiwenYang/BaryonicPSIDM}.

\begin{acknowledgments}
We thank Hai-Bo Yu, Massimo Meneghetti, and Shihong Liao for helpful discussion. The authors used generative AI tools for code-assistance and language editing during software and manuscript development. The scientific content, results, and conclusions were designed, verified, and approved by the authors.
\end{acknowledgments}

\bibliographystyle{apsrev4-2}
\bibliography{refs}

\appendix

\section{Additional Simulation Details}\label{app:simulations}

\begin{table*}[t]
\caption{Numerical settings and tidal evolution of the controlled benchmarks. The isolated evolution time is $t_{\rm iso}$. All benchmarks enter the host at $400\,\kpc$ with zero initial radial velocity and the tangential velocity $v_{\rm tan}$ listed below. Both particle species use the same softening length $\epsilon$. Final dark matter masses are measured after $2\,\gyr$ of tidal evolution. Tidal radii use the satellite mass and enclosed host mass at pericenter, with $r_s$ the initial NFW scale radius. }
\label{tab:simulation_parameters}
\centering
\small
\begin{ruledtabular}
\begin{tabular}{lccccccc}
Benchmark & $N_{\rm DM}$ & $N_\star$ & $\epsilon$ & $t_{\rm iso}$ & $v_{\rm tan}$ & $r_t/r_s$ & $M_{\rm DM,final}^{\rm SIDM}$ \\
& & & ($\kpc$) & ($\gyr$) & ($\mathrm{km/s}$) & & ($\msun$) \\
\hline
BM13.1-c & $3\times10^6$ & $10^4$ & 0.62 & 10 & 1200 & 1.96 & $4.5\times10^{12}$ \\
BM12.0 & $1.5\times10^6$ & $2\times10^4$ & 0.29 & 5 & 1200 & 2.02 & $5.2\times10^{11}$ \\
BM12.1-c & $3\times10^6$ & $10^4$ & 0.32 & 2 & 1200 & 2.84 & $6.4\times10^{11}$ \\
BM11.8 & $4\times10^6$ & $6\times10^4$ & 0.17 & 7 & 1400 & 2.86 & $3.8\times10^{11}$ \\
BM11.0-c & $2\times10^6$ & $2\times10^4$ & 0.17 & 8 & 1000 & 2.93 & $9.6\times10^{10}$ \\
BM10.8-c & $3\times10^6$ & $4\times10^4$ & 0.09 & 1.6 & 1000 & 5.67 & $5.2\times10^{10}$ \\
\end{tabular}
\end{ruledtabular}
\end{table*}

In this appendix, we provide the numerical and orbital settings of the controlled benchmarks and details of the effective-cross-section calculation. Table~\ref{tab:simulation_parameters} lists the settings for each benchmark. After isolated evolution, each subhalo is placed at $400\,\kpc$ from the host center with a purely tangential velocity of $1000$--$1400\,\mathrm{km/s}$. The fixed analytic host potential determines the subsequent orbital evolution and tidal stripping. We estimate the tidal radius using 
\begin{equation}
r_t=d\left[\frac{m}{M_{\rm host}(<d)}\right]^{1/3},
\end{equation}
where $d$ is the pericentric distance, $m$ is the satellite mass at pericenter, and $M_{\rm host}(<d)$ is the enclosed host mass.
After tidal evolution, the SIDM remnants retain at least approximately $7.8\times10^5$ dark matter particles. The adopted particle numbers and gravitational softenings accurately resolve the inner structures relevant to our lensing calculations.  

The benchmark effective cross sections in Table~\ref{tab:lensing} use the reference parameters $\sigma_0/m=10.7\,\mathrm{cm^2/g}$ and $w=505\,\mathrm{km/s}$ in Eq.~\ref{eq:dsigma}. The full angular and velocity average is \citep{Yang:2022hkm,Yang:2022zkd}
\begin{equation}
\sigma_{\rm eff}=
\frac{2\int_0^\infty {\rm d}v\int_{-1}^{1}{\rm d}\cos\theta\,
\frac{{\rm d}\sigma}{{\rm d}\cos\theta}\sin^2\theta\,v^5 f(v)}
{\int_0^\infty {\rm d}v\int_{-1}^{1}{\rm d}\cos\theta\,\sin^2\theta\,v^5 f(v)},
\label{eq:effective_cross_section}
\end{equation}
where $f(v)$ is the Maxwell--Boltzmann relative-speed distribution,
\begin{equation}
f(v)=\frac{v^2}{2\sqrt{\pi}\nu_{\rm eff}^3}
\exp\left[-\frac{v^2}{4\nu_{\rm eff}^2}\right],
\end{equation}
and $\nu_{\rm eff}=0.64V_{\max,\mathrm{NFW}}\left[1+\rho_Hr_H^2/(2\rho_sr_s^2)\right]^{1/2}$ is the characteristic one-dimensional dark matter velocity dispersion, including the deepening of the central potential by the Hernquist baryonic component.

\section{Baryonic Control of Core Formation}\label{app:Jeans Eq}
\begin{figure}[ht]
\centering
\includegraphics[width=\columnwidth]{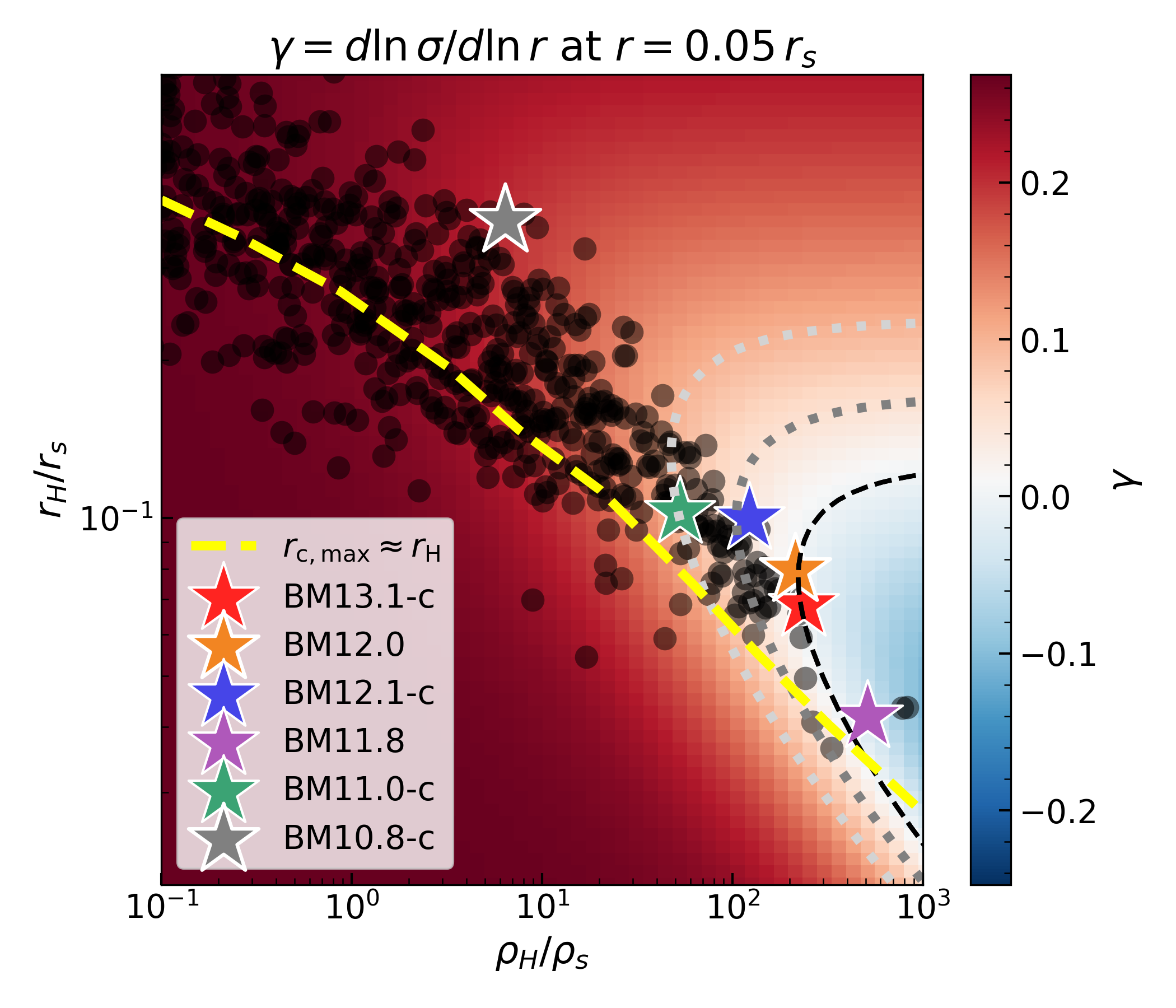}
 \caption{Velocity dispersion slope at $0.05\,r_s$ for different halo-baryon configurations. The yellow dashed line marks the parameter space where $r_{\rm c,max}=r_H$. Our benchmarks are indicated by asterisks. Substructures around the $10^{15}\,M_\odot$ host halo in TNG300-1 are shown as black scatter points. The color bar encodes the slope value at $0.05\,r_s$. Contour lines for slopes of 0, 0.05, and 0.1 are drawn in black, gray, and light gray, respectively. Benchmarks with slopes below 0.05 all skip the core formation phase.
}
 \label{fig:Jeans Eq}
\end{figure}

We use an isotropic, spherical NFW--Hernquist system to relate the stellar configuration to the initial dark matter velocity-dispersion gradient. The reference NFW dark matter profile is~\cite{Navarro:1996gj}
\begin{equation}
\rho_{\rm DM}(r) = \frac{\rho_s} {(r/r_s)(1+r/r_s)^2},    
\end{equation}
and the Hernquist stellar density is~\cite{Hernquist:1990be}
\begin{equation}
\rho_\star(r) = \frac{\rho_H} {(r/r_H)(1+r/r_H)^3}.    
\end{equation}
The enclosed stellar mass is $M_\star(<r)=2\pi\rho_Hr_H^3[r/(r+r_H)]^2$, giving $M_\star=2\pi\rho_Hr_H^3$. The quantities $\Phi_b=G\rho_Hr_H^2$ and $\Phi_h=G\rho_sr_s^2$ used in Fig.~\ref{fig:selection} are characteristic potential scales. For these untruncated profiles, the absolute central potential depths are $2\pi\Phi_b$ and $4\pi\Phi_h$, respectively.

With vanishing pressure at infinity, the isotropic Jeans equation gives
\begin{equation}\label{eq:jeans_solution}
\sigma_r^2(r) = \frac{1}{\rho_{\rm DM}(r)} \int_r^\infty \frac{G M_{\rm tot}(r')\rho_{\rm DM}(r')} {r'^2}\,{\rm d}r',
\end{equation}
where $M_{\rm tot}(r)=M_{\rm DM}(<r)+M_\star(<r)$. Define $x=r/r_s$, $\eta=r_H/r_s$, and $\delta=\rho_H/\rho_s$. These variables separate the relative stellar configuration from the absolute halo scales.

The NFW and Hernquist profiles then yield
\begin{equation}\label{eq:rho_M_dimless}
\rho_{\rm DM}(r)=\rho_s\widetilde{\rho}(x), \quad
M_{\rm tot}(r) = \rho_s r_s^3\,\tilde{M}(x,\eta,\delta),
\end{equation}
Here $\widetilde\rho(x)=[x(1+x)^2]^{-1}$ and
\begin{equation}
\begin{aligned}
\widetilde M(x;\eta,\delta)={}&4\pi\left[\ln(1+x)-\frac{x}{1+x}\right]\\
&+2\pi\delta\eta^3\frac{x^2}{(x+\eta)^2}.
\end{aligned}
\end{equation}
Substitution into Eq.~\ref{eq:jeans_solution} gives
\begin{equation}
\begin{aligned}
   & \sigma_r^2(r)=G\rho_s r_s^2\,\tilde{S}(x;\eta,\delta), \\ &\tilde{S}(x;\eta,\delta)\equiv \frac{1}{\tilde{\rho}(x)}\int_x^\infty \frac{\tilde{M}(x',\eta,\delta)\tilde{\rho}(x')}{x'^2}\,{\rm d}x'.
\end{aligned}
\label{eq:sigma_dimless}
\end{equation}
The logarithmic dispersion slope is
\begin{equation}
\gamma_\sigma(r)\equiv \frac{{\rm d}\ln\sigma_r}{{\rm d}\ln r} = \frac{1}{2}\frac{{\rm d}\ln\tilde{S}}{{\rm d}\ln x}.
\label{eq:gamma}
\end{equation}
At fixed $x$, this slope depends only on $\eta$ and $\delta$. We define $\gamma_{\sigma,0}=\gamma_\sigma(0.05r_s)$ for the
initial configuration and evaluate it numerically for each benchmark. For DM13, DM12, DM12-c, DM11, DM11-c, and DM10-c, we obtain $\gamma_{\sigma,0}=-0.0034$, $0.0025$, $0.0403$, $-0.0285$, $0.0999$, and $0.2026$, respectively. Figure~\ref{fig:Jeans Eq}
shows its dependence on the halo-baryon configuration. Small positive slopes reduce the inward heat flow that drives core expansion, while negative slopes favor outward transport. The controlled benchmarks show that core formation is suppressed for slopes below 0.1, and bypassed when the slope approaches zero or turns negative. BM11.0-c retains a shallow initial density decline, and BM10.8-c undergoes a more pronounced core-forming phase, as shown in Fig.~\ref{fig:evolution}. We therefore use $\gamma_{\sigma,0}$ as a continuous diagnostic of the initial thermal structure.

\section{Calibration and Validation of the Baryonic SIDM Model}\label{app:Recalibrated CoredDZ Profile}
\subsection{Baseline profile and gravothermal timescale}
To account for the impact of baryonic gravitational potentials on the evolution of SIDM, 
we adopt a parametric model~\cite{Yang:2024tba} that introduces a time-evolving density profile 
based on the DZ profile~\cite{Zhao:1995cp,Dekel:2017bwy,Freundlich:2020spy},  
\begin{equation}
\rho_{\text{CoredDZ}}(r) = 
\frac{f_{\rm in}(r)\rho_{\rm x}(\tau) f_{\rm out}(r)}
{\left(1 + \sqrt{r/r_x(\tau)}\right)^ {(7 - 2a)}} 
\times \frac{r_x(\tau)}{(r^k+r_c(\tau)^k)^{1/k}},
\end{equation}
where $a$ is the asymptotic inner slope of the underlying, un-cored DZ profile. It is not the central slope of the evolved profile when $r_c>0$. For a fixed reference halo and stellar configuration, the gravothermal phase is $\tau=t/t_{cb}$. The scale density $\rho_x$, scale radius $r_x$, and core radius $r_c$ evolve as
\begin{equation}
\begin{aligned}
    &\rho_x(\tau)=\rho_{x,0} \, g_\rho(\tau),\\
    &r_x(\tau)=r_{x,0} \, g_r(\tau),\\
    &r_c(\tau)=r_{s,0} \, g_c(\tau) \, \mathcal{F}_t^{2},\\
    &k = 4 \mathcal{F}_t^{1/2}, \qquad  \text{with} \quad  \mathcal{F}_t = \frac{t_{cb}}{t_c}.
\end{aligned}
\end{equation}

Here $r_{s,0}$ is the initial NFW scale radius. The initial DZ scale density and radius are
\begin{equation}
    \rho_{x,0} = (1 - a/3) \, c^3 \mu \, \bar{\rho}_h, \quad r_{x,0} = R_{\rm cut} / c,
\end{equation}
where $\mu = c^{a-3}(1 + c^{1/2})^{2(3-a)}$, $\bar{\rho}_h = M_h / (4\pi R_{\rm cut}^3/3)$, and $R_{\rm cut}$ is the virial normalization radius. The DZ concentration $c=R_{\rm cut}/r_{x,0}$ is distinct from the NFW concentration $R_{\rm vir}/r_{s,0}$.

For the conditioned benchmark comparisons, we fit $a$ and $c$ to the initial contracted dark matter profiles. This conditions the evolution test on the simulated initial state. In the population calculation, we obtain the contracted dark matter mass profile using the adiabatic-contraction prescription of Ref.~\cite{Gnedin:2004cx} and fit its DZ parameters.

We use the baryonic collapse-timescale prescription of Ref.~\cite{Yang:2024tba}. With $s=\sigma_{\rm eff}/m$, define
\begin{equation}
r_{\rm eff}=\frac{\rho_s r_s^3+\alpha\rho_Hr_H^3/2}
{\rho_s r_s^2+\alpha\rho_Hr_H^2/2},
\end{equation}
\begin{equation}
\begin{aligned}
\rho_{\rm eff}&=\frac{\rho_sr_s}{r_{\rm eff}}
+\frac{\gamma\rho_Hr_H^3}{r_{\rm eff}(r_{\rm eff}+r_H)^2},\\
v_b&=\left[4\pi G\left(\rho_sr_s^2+\frac{\alpha}{2}\rho_Hr_H^2\right)\right]^{1/2}.
\end{aligned}
\end{equation}
The baryonic and dark-matter-only (DMO) reference timescales are then
\begin{equation}
\begin{aligned}
t_{cb}&=\frac{150}{C}\frac{1}{s\rho_{\rm eff}v_b},\\
t_c&=\frac{150}{C}\frac{1}{s\rho_s(4\pi G\rho_sr_s^2)^{1/2}}.
\end{aligned}
\label{eq:baryonic_collapse_time}
\end{equation}
Here $\alpha=1.6$, $\gamma=20$, and $C=0.9$ are the reference calibration values. At the same effective cross section, $\mathcal F_t=t_{cb}/t_c$ isolates the baryonic modification of the reference timescale. The functions $g_\rho$, $g_r$, and $g_c$ retain the original DMO calibration \citep{Yang:2023jwn,Yang:2024tba}
\begin{equation}
\begin{aligned}
g_\rho(\tau)={}&2.033+0.7381\tau+7.264\tau^5-12.73\tau^7\\
&+9.915\tau^9+(1-2.033)\frac{\ln(\tau+0.001)}{\ln0.001},\\[2pt]
g_r(\tau)={}&0.7178-0.1026\tau+0.2474\tau^2-0.4079\tau^3\\
&+(1-0.7178)\frac{\ln(\tau+0.001)}{\ln0.001},\\[2pt]
g_c(\tau)={}&2.555\sqrt\tau-3.632\tau+2.131\tau^2\\
&-1.415\tau^3+0.4683\tau^4.
\end{aligned}
\end{equation}
The functions $f_{\text{in}}(r)$ and $f_{\text{out}}(r)$ are correction terms that modify the density distribution in the inner and outer regions, respectively
\begin{equation}
\begin{aligned}
&f_{\text{in}}(r) = \left(\frac{r}{r_x} + \frac{r_c}{0.4r_s}\left(\frac{\rho_{x,0}r_{x,0}}{\rho_s r_s + 0.4 \rho_H r_H}\right)^{1/(a-1)}\right)^{1-a} ,\\
&f_{\text{out}}(r) = \left(1 + \frac{r}{R_{\text{cut}}}\left(\frac{r_{x}}{r_{x,0}} - 1\right )\right)^{-1/2}.
\end{aligned}
\end{equation}

\subsection{Joint recalibration}

We recalibrate the baryonic model jointly against the simulations of Ref.~\cite{Yang:2024tba} and the six live-stellar benchmarks. The combined sample spans a broader range of halo properties, stellar mass fractions, and stellar compactness. At each stored snapshot, we fit the
full radial density profile together with the mean enclosed densities
within $0.05r_s$, $0.1r_s$, and $0.5r_s$. The snapshot loss is
\begin{equation}
\begin{aligned}
    \mathcal{L}_{\rm snap}
&=
0.20\left\langle
\left(\Delta\log_{10}\rho\right)^2
\right\rangle_{\Delta\ln r}
\\&+
\frac{0.80}{3}
\sum_{x=0.05,0.1,0.5}
\left[
\log_{10}
\frac{\bar{\rho}_{\rm model}(<xr_s)}
     {\bar{\rho}_{\rm sim}(<xr_s)}
\right]^2 ,
\end{aligned}
\end{equation}
where the radial term covers
$0.05r_s\leq r\leq R_{200}$
and is normalized in $\ln r$. The halo configurations are weighted
equally, as are the available snapshots within each configuration.
The early-time correction is
constrained using $\tau\leq0.1$, while the core-radius and radial
corrections use the full isolated evolution sequences. Three corrections address the early density evolution, the core scale, and the radial density response. The original DMO evolution functions provide the baseline, to which we apply the baryon-dependent corrections below.

\noindent\textbf{(i) Early-time regularization of \(g_\rho\).}
The original model generates an artificial rapid density increase near $\tau \simeq 0$ in some compact baryonic configurations. This increase is absent in the simulations. The phenomenon primarily arises from the mismatch between the DMO scale density evolution function and the strong baryonic initial state. We apply early-time regularization solely to $g_\rho$ and keep $g_r$ unchanged.
\begin{equation}
g_{\rho,\rm new}(\tau) = g_{\rho,\rm DMO}(\tau) \exp[-\Delta_{\rm early}(\tau)],    
\end{equation}
where
\begin{equation}
\begin{aligned}
\Delta_{\rm early}(\tau)={}&G_JU_\Phi A_{\rm early}(\eta)\\
&\times\left[1-\exp\left(-\left(\frac{\tau}{0.00799557}\right)^{0.78032133}\right)\right],\\
A_{\rm early}(\eta)={}&0.27194951+\frac{2.54505890}{1+(\eta/0.055)^6}.
\end{aligned}
\end{equation}
We further define
\begin{equation}
\eta_\Phi = \frac{\rho_Hr_H^2}{\rho_s r_s^2}, \qquad U_\Phi = \frac{\eta_\Phi} {\eta_\Phi+0.87767347}, \qquad \eta=\frac{r_H}{r_s}.    
\end{equation}
where $\eta_\Phi$ measures the baryonic potential depth relative to the dark matter potential depth, and $\eta$ describes the compactness of the baryonic distribution. This correction reaches a significant magnitude only when the baryonic potential is highly significant and the baryonic distribution is highly compact. It saturates rapidly after the early phase. Therefore, this correction does not persistently alter the late-time evolution slope.
To avoid unnecessary suppression for configurations without the anomalous early increase, we further introduce an early-response gate.
\begin{equation}
G_J = \frac{1} {1+\exp[-(J-\log_{10}(1.20))/0.00055]},    
\end{equation}
where
\begin{equation}
J = \max_{R=0.1r_s,\,0.5r_s} \left[ \log_{10} \frac{ \bar{\rho}_{\rm raw}(<R,\tau_p) }{ \bar{\rho}_{\rm raw}(<R,0) } \right], \qquad \tau_p=0.08.    
\end{equation}
Here $\bar{\rho}_{\rm raw}(<R)$ is the mean enclosed density predicted by the baseline Cored-DZ model. The gate $G_J$ approaches zero for early increases below approximately $20\%$ and activates the correction for larger increases.

\noindent\textbf{(ii) Velocity-dispersion-slope-dependent correction to $r_c$.} For configurations with a sufficiently positive inner dispersion slope, the original $\mathcal F_t^2$ prescription overly suppresses $r_c$. Using $\gamma_{\sigma,0}=\gamma_\sigma(0.05r_s)$ from Sec.~\ref{app:Jeans Eq}, we introduce
\begin{equation}
S_\sigma = \frac{1} { 1+\exp\left[ -\dfrac{\gamma_{\sigma,0}-0.099} {0.00393036} \right] },   
\end{equation}
The corrected core radius is given by
\begin{equation}
r_c(\tau) = r_{s,0}g_c(\tau) \left[ \mathcal F_t^2 + 0.15030104S_\sigma (1-\mathcal F_t^2) \right].
\end{equation}
For large positive $\gamma_{\sigma,0}$, $S_\sigma$ approaches unity and partially restores the core scale. For small or negative slopes, $S_\sigma$ approaches zero and the original $\mathcal F_t^2$ scaling is recovered.

\noindent\textbf{(iii) Mass-conserving radial response.} Strong baryonic potential configurations with suppressed cores still exhibit a growing radial residual after the corrections to the early density and core radius. The simulated inner density typically grows faster than the baseline model. The larger radii require weaker or even opposite compensation. We denote the Cored-DZ profile after early and core corrections as \(\rho_{\rm base}\). We introduce a mass-conserving radial response.
\begin{equation}
\begin{aligned}
    \rho_{\rm final}(r,\tau) &= {\cal N}(\tau)\rho_{\rm base}(r,\tau)\\ &\times\exp\left\{ S(\tau) 
    \left[ K_{\rm in}(r) - C_{\rm M}(\tau)K_{\rm out}(r) \right] \right\},
\end{aligned}
\end{equation}
where response strength $S(\tau)$ is defined as
\begin{equation}
\begin{aligned}
S(\tau)={}&1.2023\,U_\Phi(1-S_\sigma)\\
&\times\left[1-\exp\left(-\left(\frac{\tau}{0.37366}\right)^{1.7410}\right)\right].
\end{aligned}
\end{equation}
The factor $1-S_\sigma$ restricts this radial correction mainly to configurations with suppressed core formation. This restriction avoids significant changes to halos already well-described by the core radius correction.

The inner enhancement kernel and the outer compensation kernel are respectively given by

\begin{equation}
\begin{aligned}
&K_{\rm in}(r) = \exp\left( -\frac{r}{0.20964676r_{s,0}} \right),\\  
K_{\rm out}(r)&=\left(\frac{r}{0.31328434r_{s,0}}\right)^2\\
&\quad\times\exp\left[-\left(\frac{r}{0.31328434r_{s,0}}\right)^2\right].
\end{aligned}
\end{equation}
The compensation coefficient is
\begin{equation}
C_{\rm M}(\tau) = \frac{ \displaystyle \int_0^{R_{\rm cut}} \rho_{\rm base}(r,\tau) K_{\rm in}(r)r^2\,{\rm d}r }{ \displaystyle \int_0^{R_{\rm cut}} \rho_{\rm base}(r,\tau) K_{\rm out}(r)r^2\,{\rm d}r }
\end{equation}
This coefficient zeroes the linear mass moment of \(K_{\rm in}-C_{\rm M}K_{\rm out}\). The final correction adopts an exponential form. We further use
{\small
\begin{equation}
{\cal N}(\tau) = \frac{ \displaystyle \int_0^{R_{\rm cut}} \rho_{\rm base}(r,\tau)r^2\,{\rm d}r }{ \displaystyle \int_0^{R_{\rm cut}} \rho_{\rm base}(r,\tau) e^{S(\tau) [K_{\rm in}(r)-C_{\rm M}(\tau)K_{\rm out}(r)] } r^2\,{\rm d}r }
\end{equation}}
for exact normalization. This normalization guarantees
\begin{equation}
M_{\rm final}(<R_{\rm cut},\tau) = M_{\rm base}(<R_{\rm cut},\tau)
\end{equation}

The three corrections act on distinct features of the evolution. In the no-baryon limit, $U_\Phi=0$ and $\mathcal F_t=1$, so the early and radial corrections vanish and the core prescription recovers the baseline. At $\tau=0$, the corrections preserve the adopted initial DZ profile. Figure~\ref{fig:evolution} shows the isolated density evolution obtained with this joint calibration.

\subsection{Integral approach and tidal tests}

\begin{figure}[htbp]
	\centering
\includegraphics[width=\columnwidth]{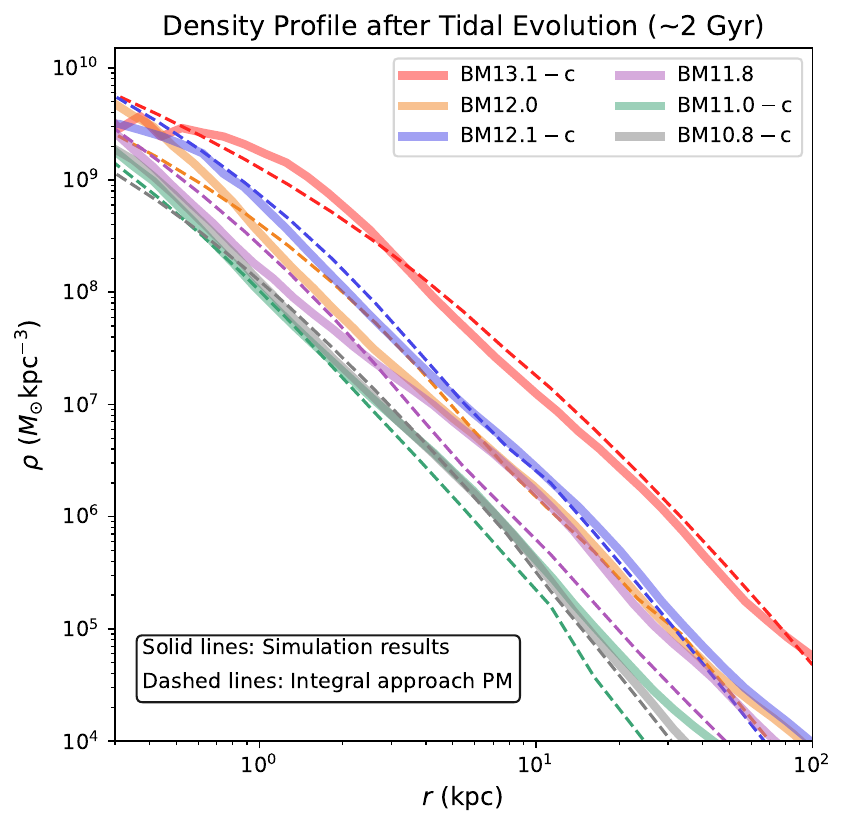}
	\caption{Dark matter density profiles of the six benchmarks after approximately
$2\,\mathrm{Gyr}$ of tidal evolution.
Thick solid curves show the simulation results.Thin dashed curves show the integral-approach parametric predictions.}
	\label{fig:tidal_profile_comparison}
\end{figure}

\begin{figure*}[htbp]
	\centering
\includegraphics[width=\columnwidth]{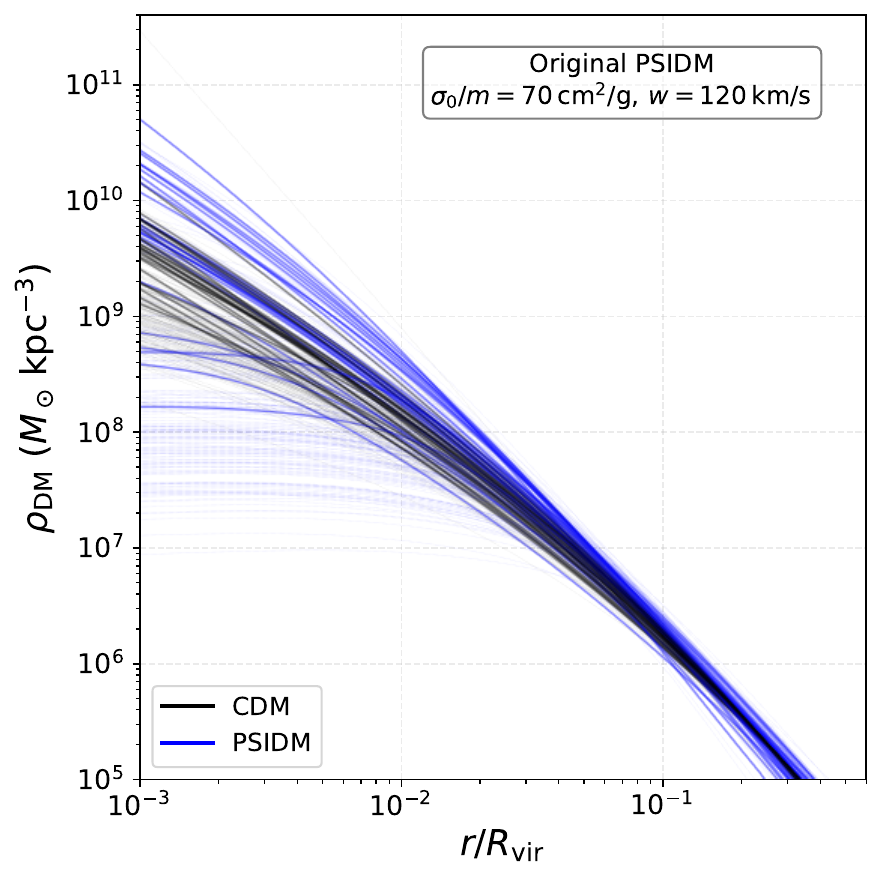}
\includegraphics[width=\columnwidth]{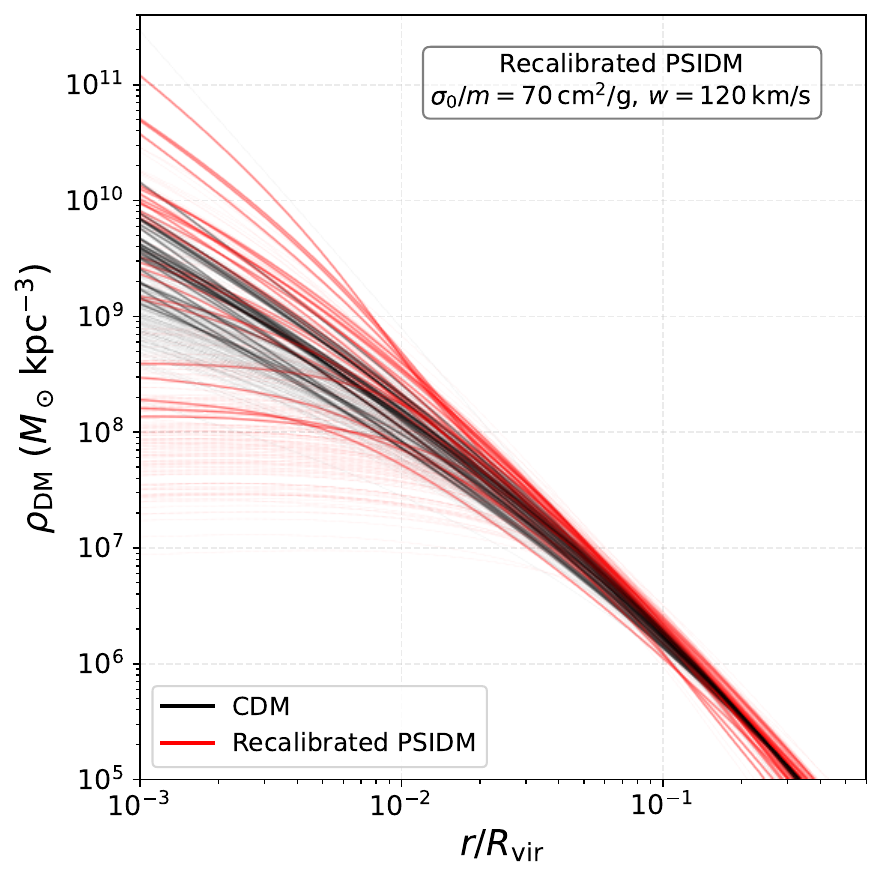}
\caption{Effect of baryonic PSIDM recalibration on cluster subhalo density profiles. The original model of Ref.~\cite{Yang:2024tba} (left, blue) and the recalibrated model (right, red) are applied to subhalos of the most massive TNG300-1 cluster, with matched CDM profiles shown in black. Both calculations use $\sigma_0/m=70\,\mathrm{cm^2/g}$ and $w=120\,\mathrm{km/s}$. Solid curves highlight the same subhalos in both panels, selected by $0.5''<\theta_{\rm E,eff}<3''$ in the recalibrated SIDM model. Faint dotted curves show the remaining subhalos with masses above $10^{10}\,\msun$. Radii are normalized by the halo virial radius $R_{\rm vir}$.}
	\label{fig:PM_density_profiles_w120_sigma0_70}
\end{figure*}

The integral approach adds gravothermal structural changes to a CDM evolution history \citep{Yang:2023jwn,Yang:2024uqb,Yang:2024tba}. We set $\tau=0$ at the halo formation time, defined as the first time its main progenitor reaches half of its mass at the selected snapshot. Taking $t$ as the elapsed time since formation, the gravothermal phase is
\begin{equation}
\tau(t)=\int_0^t\frac{{\rm d}t'}{t_{cb}[s(t'),\mathbf p(t')]},
\label{eq:integral_phase}
\end{equation}
where $\mathbf p(t)=(\rho_s,r_s,\rho_H,r_H)(t)$ and $s(t)=\sigma_{\rm eff}(t)/m$. The ratio $t/t_{cb}$ applies only when the reference timescale is constant. The integral mapping can be written for either $X=V_{\rm max}$ or $R_{\rm max}$ as
\begin{equation}
X_{\rm SIDM}(t)=X_{\rm CDM}(t)+\Delta X(t),
\end{equation}
\begin{equation}
\Delta X(t)=\int_0^t\frac{{\rm d}t'}{t_{cb}(t')}
\left.\frac{\partial X_{\rm model}(\tau;\mathbf p)}{\partial\tau}
\right|_{\tau(t'),\mathbf p(t')}.
\label{eq:integral_mapping}
\end{equation}
The derivative describes gravothermal evolution at fixed instantaneous reference parameters. Evaluating it along the CDM history incorporates the changing halo structure during accretion and mass loss. Here $V_{\rm max}$ and $R_{\rm max}$ refer to the dark matter circular-velocity curve.

For an NFW reference progenitor, its scale parameters satisfy
\begin{equation}
r_s=\frac{R_{\rm max,0}}{2.1626},\qquad
\rho_s=\frac{1}{G}\left(\frac{V_{\rm max,0}}{1.648r_s}\right)^2.
\label{eq:nfw_reconstruction}
\end{equation}
The baryonic profile is reconstructed from the reference halo, stellar parameters, and accumulated phase using the recalibrated model. 
For profile evaluation, we use $\tau_{\rm eval}=\min(\tau,1.1)$, following Ref.~\cite{Yang:2024tba}.
For the population calculation, we adopt the TNG300-1 snapshot closest to the lens redshift, $z_l=0.439$. The stellar distributions are held identical to CDM in the population application. No additional tidal-truncation prescription is imposed on the predicted SIDM profile.

Figure~\ref{fig:tidal_profile_comparison} compares the integral predictions with the simulated dark matter profiles after $2\,\gyr$ in the host potential. It tests the model response to mass loss in the presence of stars, complementing the isolated evolution shown in Fig.~\ref{fig:evolution}. A stochastic extension that incorporates SIDM-induced structural changes along halo assembly histories is discussed in Ref.~\cite{Yang:2026wln}.

Figure~\ref{fig:PM_density_profiles_w120_sigma0_70} compares cluster subhalo density profiles predicted by the original baryonic PSIDM model (left, blue) and its recalibrated version (right, red), with matched CDM profiles shown in black. Both calculations adopt the velocity-dependent SIDM model with $\sigma_0/m=70\,\mathrm{cm^2/g}$ and $w=120\,\mathrm{km/s}$, using identical halo histories and stellar distributions. Solid curves highlight the same subhalos in both panels, selected by $0.5''<\theta_{\rm E,eff}<3''$ in the recalibrated model. Faint dotted curves show the remaining subhalos with masses above $10^{10}\,\msun$. The recalibration changes the strength and radial extent of the inner density enhancement while leaving the outer profiles broadly similar. In both versions, a majority of the lensing-selected subhalos have higher inner dark matter densities than their CDM counterparts. 

\section{Strong-Lensing Calculations}\label{app:GGSL_benchmarks}
\subsection{Projection and lensing geometry}

For a spherically averaged benchmark, the total density is $\rho_{\rm tot}(r)=\rho_{\rm DM}(r)+\rho_\star(r)$. Its projected density and enclosed projected mass are
\begin{equation}
\Sigma(R)=2\int_R^{R_{\rm out}}
\frac{\rho_{\rm tot}(r)r\,{\rm d}r}{\sqrt{r^2-R^2}},
\end{equation}
\begin{equation}
M_{\rm 2D}(<R)=2\pi\int_0^R\Sigma(R')R'\,{\rm d}R'.
\end{equation}
Here $R_{\rm out}$ is the adopted outer integration boundary. The convergence is $\kappa=\Sigma/\Sigma_{\rm cr}$, where
\begin{equation}
\Sigma_{\rm cr}=\frac{c_{\rm light}^2}{4\pi G}
\frac{D_s}{D_lD_{ls}}.
\end{equation}
The angular-diameter distances $D_l$, $D_s$, and $D_{ls}$ connect the observer, lens, and source, and $c_{\rm light}$ is the speed of light. For a circular lens, the reduced deflection is
\begin{equation}
\alpha(\theta)=\bar\kappa(<\theta)\theta
=\frac{M_{\rm 2D}(<D_l\theta)}{\pi D_l^2\Sigma_{\rm cr}\theta}.
\label{eq:reduced_deflection}
\end{equation}
Angles in these expressions are in radians. Numerical results may be converted to arcseconds after evaluating the deflection.

For the benchmark critical curves and caustics, we use $z_l=0.439$, $z_s=3$, and a projected cluster-centric distance of $350\,\kpc$. CDM and SIDM receive the same smooth-host contribution.  We evaluate the convergence on a $2000\times2000$
grid and obtain the deflection and shear by solving $\nabla_{\boldsymbol{\theta}}^2\psi=2\kappa$ with a fast Fourier
transform. The tangential critical curve satisfies $\lambda_t=1-\kappa-|\gamma|=0$. Mapping it through $\boldsymbol\beta=\boldsymbol\theta-\boldsymbol\alpha(\boldsymbol\theta)$ gives the tangential caustic. Its source-plane area is $\sigma_T$, while the image-plane critical area defines $\theta_{\rm E,eff}=\sqrt{A_{\rm crit}/\pi}$.

\subsection{Benchmark lensing results}

\begin{figure*}[htbp]
	\centering
	\includegraphics[width=16.5cm,height=5.2cm]{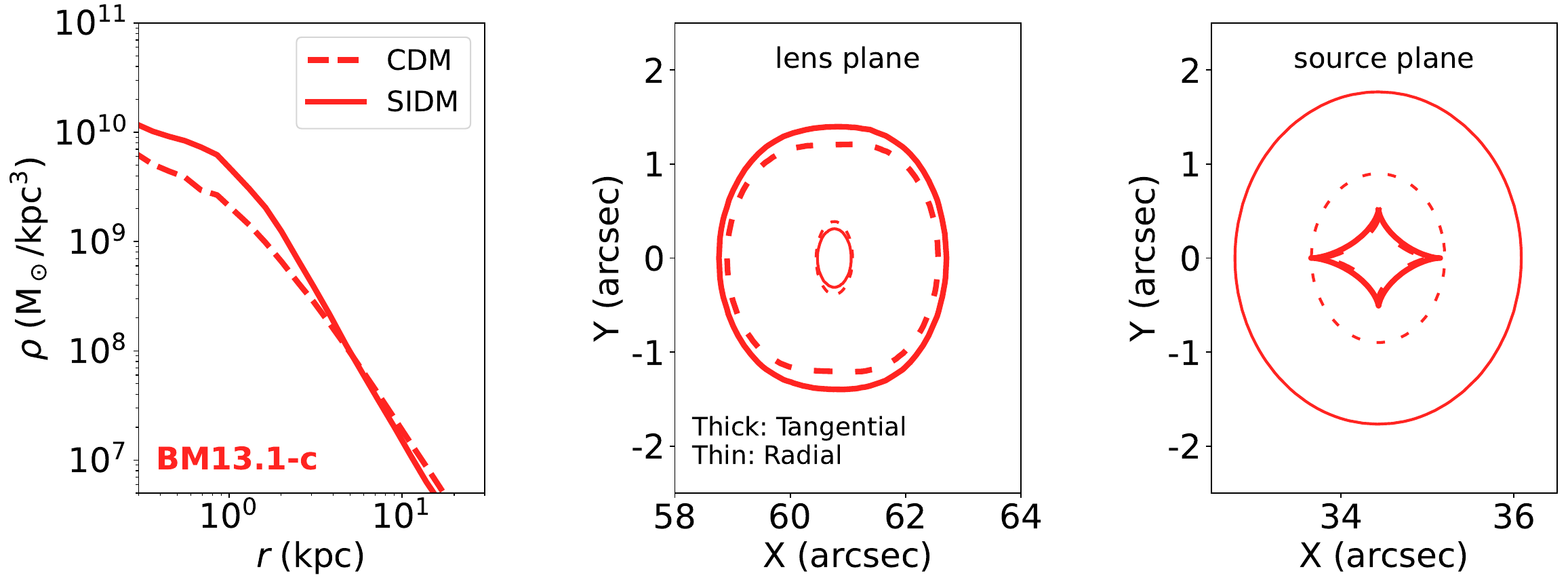}\\[0.1cm]
	\includegraphics[width=16.5cm,height=5.2cm]{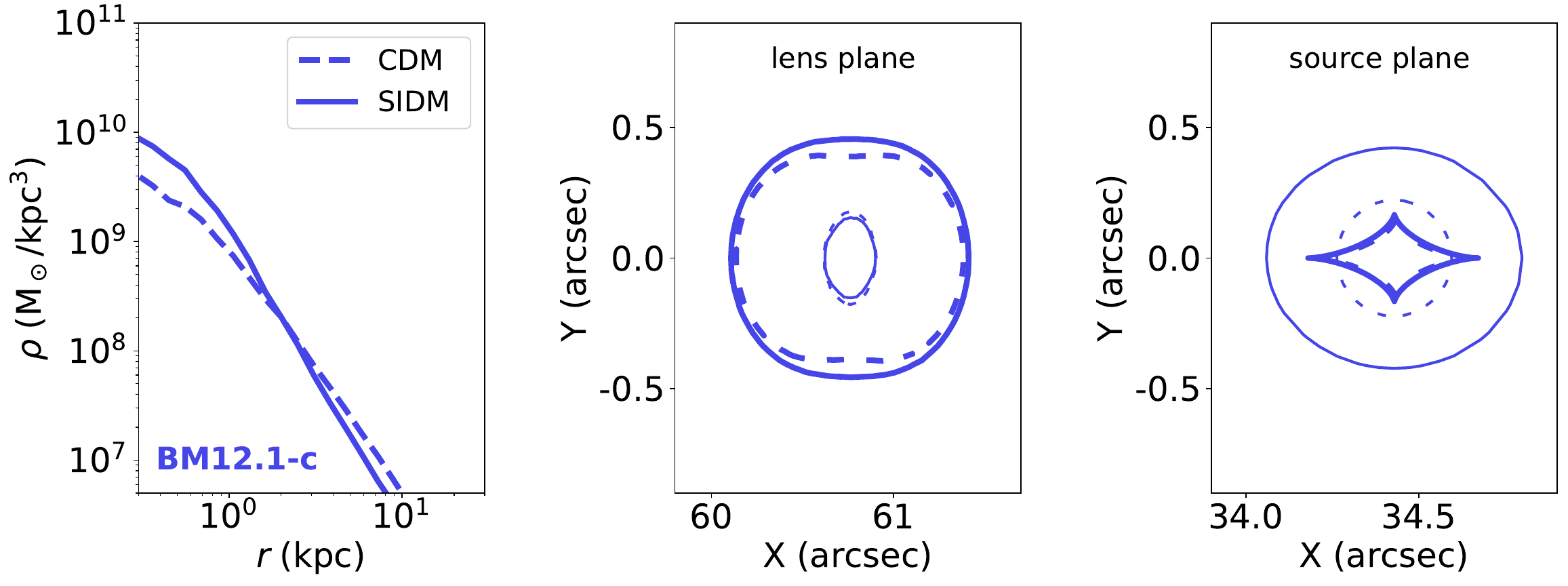}\\[0.1cm]
	\includegraphics[width=16.5cm,height=5.2cm]{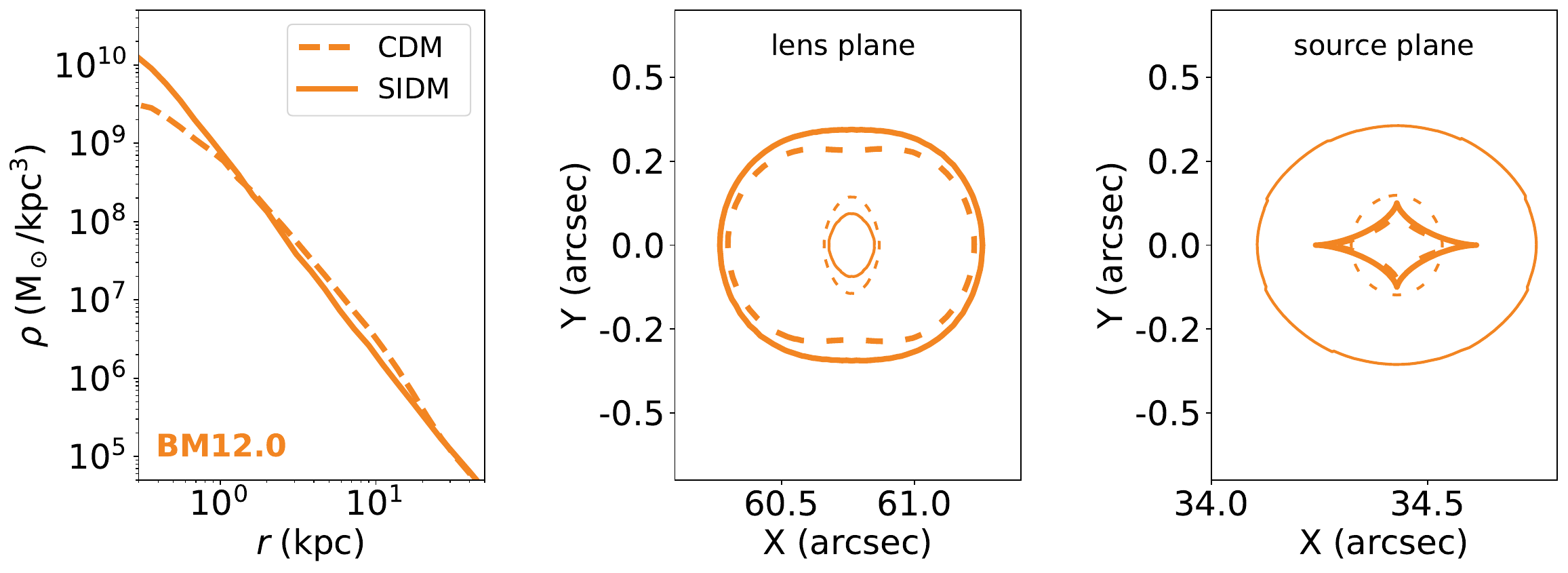}
	\caption{Density profiles, lens-plane critical curves, and source-plane caustic curves for BM13.1-c (top), BM12.1-c (middle), and BM12.0 (bottom). Solid lines represent SIDM models and dashed lines CDM models.}
	\label{Fig:BM_appendix_1}
\end{figure*}

\begin{figure*}[htbp]
	\centering
	\includegraphics[width=16.5cm,height=5.2cm]{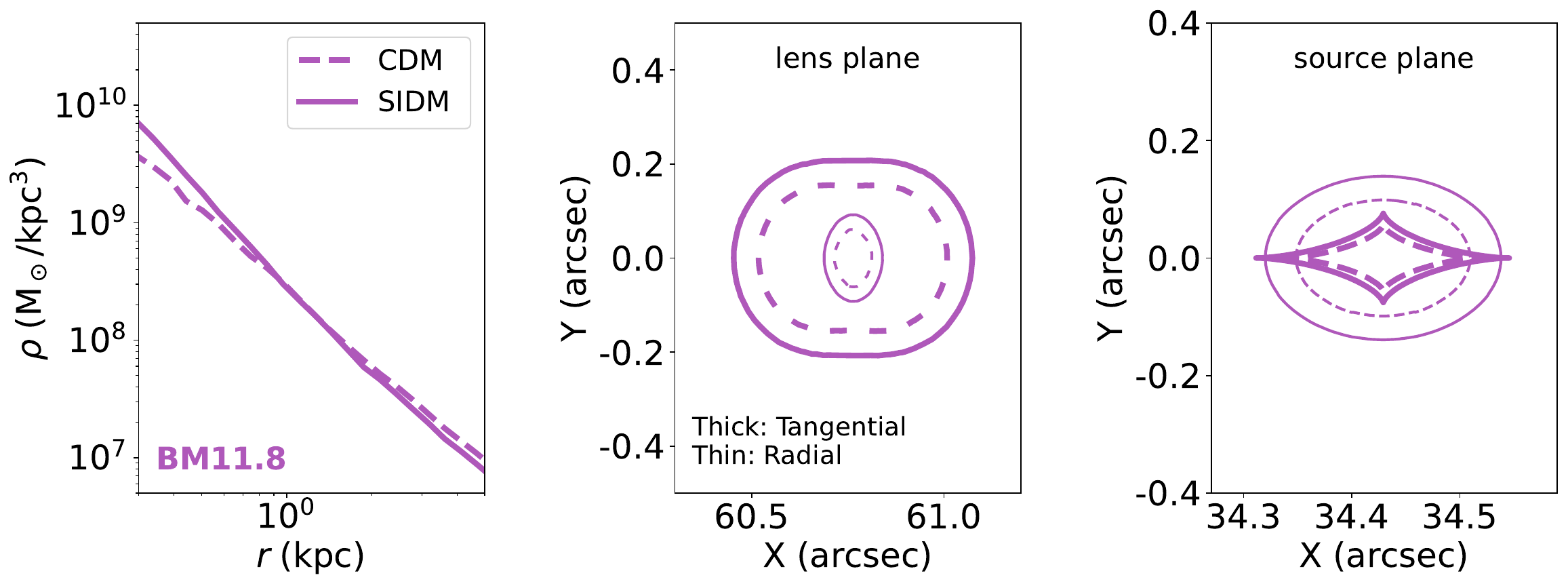}\\[0.1cm]
	\includegraphics[width=16.5cm,height=5.2cm]{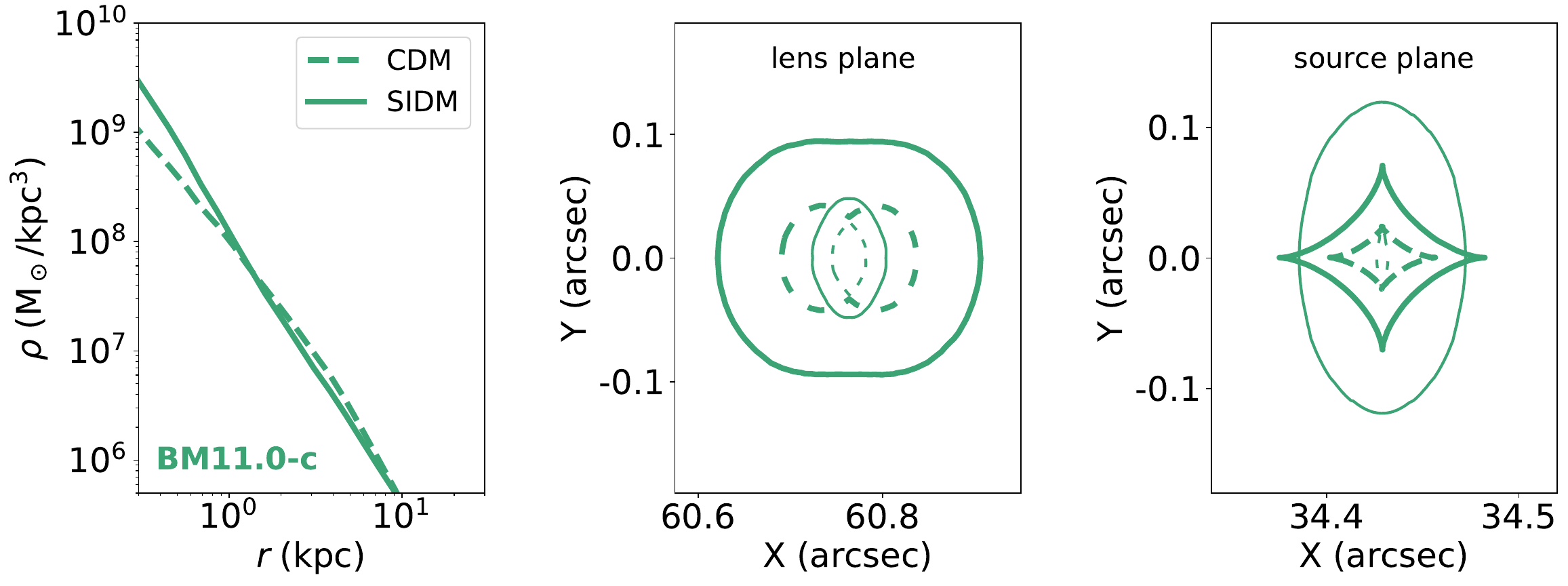}\\[0.1cm]
	\includegraphics[width=16.5cm,height=5.2cm]{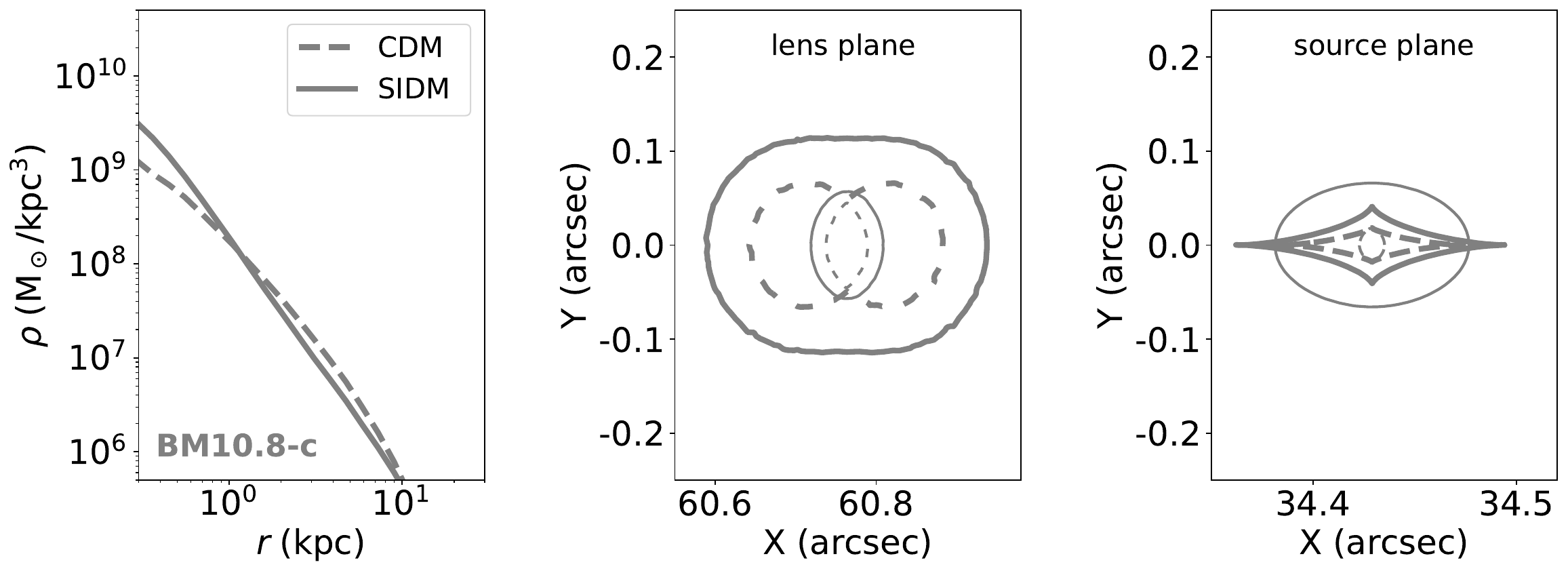}
	\caption{Density profiles, lens-plane critical curves, and source-plane caustic curves for BM11.8 (top), BM11.0-c (middle), and BM10.8-c (bottom). The SIDM models consistently exhibit a higher central density and larger critical and caustic areas compared to the CDM models.}
	\label{Fig:BM_appendix_2}
\end{figure*}

Figures~\ref{Fig:BM_appendix_1} and \ref{Fig:BM_appendix_2} give the complete paired benchmark results. Solid and dashed curves distinguish SIDM and CDM, while thin and thick curves identify radial and tangential branches in the lensing panels. The tangential caustic-area ratios are $1.89$, $1.91$, and $1.29$ for BM13.1-c, BM12.1-c, and BM12.0, respectively. The remaining benchmarks give $1.65$, $5$, and $3.17$ for BM11.8, BM11.0-c, and BM10.8-c. Table~\ref{tab:lensing} reports these results at rounded precision.

The strongest inner contraction need not produce the largest lensing enhancement because the relevant projected mass depends on the extent of contraction relative to the critical curve. Several benchmarks have effective Einstein radii below the $0.5''$ threshold of the population calculation. Their role is to test the structural and lensing response over a range of halo properties, while the population calculation determines which subhalos contribute above the adopted threshold.

\subsection{Cluster populations}

The observed small-scale lenses mostly have effective Einstein radii $0.5''\lesssim\theta_{\rm E,eff}\lesssim3''$ and are associated with subhalos of total masses approximately $10^{11}$--$10^{12}\,\msun$~\citep{Meneghetti:2020yif,Meneghetti:2022apr,Meneghetti:2023fug}.
We select the three most massive TNG300-1 clusters at the snapshot nearest $z_l=0.439$ and apply the calibrated model to their resolved subhalo populations. The host and subhalo properties are evaluated at this snapshot, which also sets the endpoint of the integral evolution. 

We match subhalos between the DMO and full-physics simulations. The DMO particle mass is $7.0\times10^7\,\msun$, giving approximately 1400 particles per $10^{11}\,\msun$ subhalo and providing the resolved CDM assembly histories used in the mapping~\citep{Nelson:2017cxy,Pillepich:2017fcc}. We reconstruct the reference NFW parameters from the DMO $V_{\max}$ and $R_{\max}$ histories and obtain the Hernquist stellar parameters from the matched full-physics halos. SIDM enters as controlled structural modifications along these CDM histories. The paired predictions retain the same stellar distributions, subhalo catalog, host potential, and lensing geometry.

Eleven views are used for each host, comprising the original orientation and five rotations in $18^\circ$ increments about each of two orthogonal axes. For model $A\in\{\mathrm{CDM},\mathrm{SIDM}\}$ and viewing direction $j$, define
\begin{equation}
\sigma_{{\rm GGSL},j}^{A}=\sum_{i\in\mathcal L_j^A}\sigma_{T,i,j}^{A},
\end{equation}
where $\mathcal L_j^A$ denotes the selected subhalo lenses within the $200''\times200''$ field, retaining only candidates with $0.5''<\theta_{\rm E,eff}\leq3''$. The reported enhancement is
\begin{equation}
\mathcal R_{\rm GGSL}=
\frac{\sum_{j=1}^{11}\sigma_{{\rm GGSL},j}^{\rm SIDM}}
{\sum_{j=1}^{11}\sigma_{{\rm GGSL},j}^{\rm CDM}}.
\label{eq:population_ratio}
\end{equation}
This is the ratio of separately averaged total cross sections, not the average of individual-lens or per-view ratios. Table~\ref{tab:ggsl} gives these ratios for both scattering models in order of decreasing host mass. Figure~\ref{fig:population} instead shows individual matched-lens ratios, plotted against the SIDM effective Einstein radius. The viewing directions sample orientation dependence within three hosts.

\begin{table*}[t]
\caption{SIDM-to-CDM GGSL cross-section ratios for the three most massive TNG300-1 clusters at the snapshot nearest $z_l=0.439$, predicted using the recalibrated baryonic parametric SIDM model. Here $M_h$ is the host cluster mass and $M_b$ is the stellar mass of its central galaxy.}
\label{tab:ggsl}
\centering
\small
\begin{ruledtabular}
\begin{tabular}{lccccc}
Host & $M_h$ & $M_b$ & $w$ & $\sigma_0/m$ &
$\langle\sigma_{\rm GGSL}^{\rm SIDM}\rangle/
 \langle\sigma_{\rm GGSL}^{\rm CDM}\rangle$ \\
& ($M_\odot$) & ($M_\odot$) &
($\mathrm{km\,s^{-1}}$) & ($\mathrm{cm^2\,g^{-1}}$) & \\
\hline
Halo 1 & $1.36\times10^{15}$ & $4.46\times10^{12}$ & 500 & 20 & 1.51 \\
       &                   &                    & 120 & 70 & 1.45 \\
\hline
Halo 2 & $8.89\times10^{14}$ & $2.84\times10^{12}$ & 500 & 20 & 1.43 \\
       &                   &                    & 120 & 70 & 2.65 \\
\hline
Halo 3 & $5.32\times10^{14}$ & $2.41\times10^{12}$ & 500 & 20 & 2.34 \\
       &                   &                    & 120 & 70 & 1.67 \\
\end{tabular}
\end{ruledtabular}
\end{table*}

The GGSL probability scales approximately with the square of the host Einstein radius, making host selection important~\citep{Roche:2026jan}. Our matched CDM--SIDM calculations fix the cluster sample and lensing environment, helping isolate the SIDM enhancement from sample-selection biases. 

Existing hydrodynamical cluster simulations have investigated both velocity-independent~\citep{Robertson:2018anx,Harvey:2025paa,Lee:2026otu} and velocity-dependent SIDM~\citep{Robertson:2018anx,Ragagnin:2024deh}. Isolated-halo tests report significant scatter in collapse times below approximately $10^5$ particles~\citep{Mace:2024uze}, corresponding to a reference particle mass of $10^6\,\msun$ for a $10^{11}\,\msun$ dark matter subhalo. Dianoga SIDM and TNGC-SIDM use dark matter particle masses of $1.15\times10^9\,\msun$ and $6.1\times10^7\,\msun$, respectively~\citep{Ragagnin:2024deh,Lee:2026otu}, yielding approximately 90 and 1600 particles per such subhalo. Reaching the reference target requires mass-resolution improvements of approximately $10^3$ and 60. Our model provides lensing predictions for tests that resolve this internal evolution.

\subsection{Luminous-satellite illustration}

For Fig.~\ref{fig:perturber}, we evaluate Eq.~\ref{eq:reduced_deflection} for the total simulated CDM and SIDM profiles of BM11.8. The comparison uses the stellar contribution from each paired simulation. The published G4 lens models provide a reference for the deflection scale of a luminous satellite \citep{Vegetti:2010wa}.

We express the Pseudo-Jaffe model entirely in angular coordinates. With angular lens strength $b_{\rm sub}$ and angular truncation scale $\theta_t$, its convergence is
\begin{equation}
\kappa_{\rm PJ}(\theta)=\frac{b_{\rm sub}}{2}
\left[\frac{1}{\theta}-\frac{1}{\sqrt{\theta^2+\theta_t^2}}\right],
\end{equation}
and its reduced deflection is
\begin{equation}
\alpha_{\rm PJ}(\theta)=\frac{b_{\rm sub}}{\theta}
\left[\theta+\theta_t-\sqrt{\theta^2+\theta_t^2}\right].
\label{eq:pj_deflection}
\end{equation}
The corresponding total projected mass normalization is $M_{\rm PJ}=\pi\Sigma_{\rm cr}D_l^2b_{\rm sub}\theta_t$. The PL$+$2SIS$+$PJ fit uses $M_{\rm PJ}=2.78\times10^{10}\,\msun$ and $\theta_t=0.68''$, while the SIE$+$PJ fit uses $2.75\times10^{10}\,\msun$ and $0.81''$ \citep{Vegetti:2010wa}. These are two lens-model reference curves, not an uncertainty band. The angular truncation scale is a Pseudo-Jaffe profile parameter tied to the adopted lens model and is distinct from a dynamical tidal-radius estimate based on the satellite orbit.

The published system has $z_l=0.422$ and $z_s=2.001$, with $H_0=73\,\mathrm{km\,s^{-1}\,Mpc^{-1}}$, $\Omega_m=0.25$, and $\Omega_\Lambda=0.75$ \citep{Vegetti:2010wa}. These values give $1''\simeq5.40\,\kpc$. The positions of G4 and the dominant lens galaxy G1 imply a projected separation of $4.16''\simeq22.5\,\kpc$.
The simulated deflection profiles are evaluated at the observed lens and source redshifts.

The tidal scaling in the main text compares environments through the approximation $r_t\propto d(m/M_{\rm host})^{1/3}$. At fixed satellite mass and tidal radius, replacing a $10^{15}\,\msun$ cluster with a $10^{13}\,\msun$ group maps $d=320\,\kpc$ to $320(10^{13}/10^{15})^{1/3}\simeq69\,\kpc$. This is an order-of-magnitude comparison using total host masses. 
The G4 projected separation allows a larger three-dimensional distance but does not determine its orbit or stripping history.

\end{document}